\documentclass[runningheads,a4paper]{llncs}

\usepackage[a4paper,margin=1in]{geometry}

\usepackage[utf8]{inputenc}
\usepackage[T1]{fontenc}
\usepackage{lmodern}
\usepackage{textcomp}
\usepackage[normalem]{ulem}

\usepackage{amssymb}
\usepackage{amsmath}
\usepackage{graphicx}

\usepackage{url}
\usepackage{booktabs}
\usepackage{longtable}
\usepackage{array}
\usepackage{calc}

\usepackage{booktabs}
\usepackage{tabularx}
\usepackage{array}

\newcommand{\keywords}[1]{\par\addvspace\baselineskip
\noindent\keywordname\enspace\ignorespaces#1}

\begin{document}

\mainmatter

\title{Thermodynamic principles of emerging cryopreservation technologies}
\titlerunning{Thermodynamic principles of emerging cryopreservation technologies}

\author{Matthew J. Powell-Palm\inst{1,2,3,4} \and Anthony N. Consiglio\inst{4,5}}
\authorrunning{Powell-Palm and Consiglio}

\institute{J. Mike Walker '66 Department of Mechanical Engineering, Texas A\&M University, College Station, TX, USA \and
Department of Materials Science \& Engineering, Texas A\&M University, College Station, TX, USA \and
Department of Biomedical Engineering, Texas A\&M University, College Station, TX, USA \and
BioChoric Inc., Bozeman, MT, USA \and
Department of Mechanical Engineering, University of California, Berkeley, Berkeley, CA, USA,\\
\email{powellpalm@tamu.edu}, \\
\email{aconsiglio4@berkeley.edu}}

\toctitle{Thermodynamic principles of emerging cryopreservation technologies}
\tocauthor{Powell-Palm and Consiglio}
\maketitle

\begin{abstract}
Modern cryopreservation exists at the convergence of diverse disciplines---materials science, physical chemistry, mechanical engineering, biological engineering, etc.---and emerging technologies often draw from many of these disciplines simultaneously. Thermodynamics, as one of the foundational theories underlying both physical and biological science, provides a framework through which to understand these interdisciplinary technologies, yet the full kit of requisite thermodynamic tools is not housed within any one discipline. This Chapter aims to articulate a foundational thermodynamic approach to the description, interrogation, and design of modern cryopreservation technologies, and to review the state of the art in emerging cryopreservation technologies through the lens of this approach. We focus in particular on the management of phase change across equilibrium-driven techniques (e.g. liquidus tracking, partial freezing, isochoric freezing), kinetics-driven techniques (e.g. supercooling, ice seeding), and transport-driven techniques (e.g. directional freezing, droplet approaches), and we hope to equip the reader with a self-consistent theoretical toolkit that enables meaningful comparison of these techniques from a thermodynamic perspective.
\end{abstract}

\section*{Preprint acknowledgment}
This is a preprint of the following chapter: Matthew J. Powell-Palm and Anthony N. Consiglio, Thermodynamic principles of emerging cryopreservation technologies, published in Cryopreservation and Freeze-Drying Protocols, edited by Willem F. Wolkers and Harriëtte Oldenhof, 2026, Springer. It is the version of the author’s manuscript prior to acceptance for publication and has not
undergone editorial and/or peer review on behalf of the Publisher (where applicable).

\newpage

\section{Introduction}
Effective cryopreservation of biological matter---be that matter a cell, a tissue, an organ, or an organism---hinges upon the uniquely
complicated interplay between the thermodynamics of aqueous solutions, the nucleation and growth kinetics of ice and other solid phases, and the transport of heat and mass through spatially and chemically heterogeneous media. As such, mathematical description, interrogation, and design of cryopreservation processes require a uniquely varied blend of physical theories, drawn from disciplines ranging from physical chemistry and atmospheric science to mechanical engineering and materials science. Classical geometric (or Gibbsian) thermodynamics can provide a unifying framework to connect these disparate bases of knowledge, but reconciliation of different schools of thermodynamic practice in the context of cryopreservation is required.

To that end, in this Chapter, we will present a thermodynamic
description of the physical aspects of cryopreservation, enabling
self-consistent interrogation of the physical bases of various emergent
cryopreservation technologies. In the first Section, we will establish
this description in additive sub-sections focusing on solution
thermodynamics, ice nucleation \& growth kinetics, and heat and mass
transport. In each sub-section, we will aim to equip the user firstly
with high-level physical insight applicable to emergent cryopreservation
technologies, and secondly with basic predictive power over the same. In
the second Section, we will review various such technologies within the
context of our unified physical framework, aiming to facilitate
interrogation of how the underlying thermodynamic premise of each
technology may resemble or differ from the others.

For consistency with our theoretical approach, we will group these
emergent technologies by the theoretical aspect---equilibrium
thermodynamics, kinetics, or transport---that most principally drives
them. For example, liquidus tracking, partial freezing \cite{r1},
and isochoric freezing \cite{r2} all rely upon attainment of a
desired state of thermodynamic equilibrium, wherein ice may be either
entirely absent (liquidus tracking) or allowed to an extent prescribed
by the liquidus equilibria of the solution (partial freezing and
isochoric freezing). Supercooling \cite{r3} and ice seeding, by
comparison, rely upon attainment of a desired nucleation rate
(supercooling) or ice growth rate (seeding), which, while of course
influenced by the underlying phase equilibria, are ultimately kinetic
factors. Likewise, various other techniques rely upon attainment of a
desired temperature profile \emph{en route} to the final state of
preservation, be it a profile in time (as in droplet and mesh
technologies that seek to amplify cooling rates) or in space (as in
directional freezing). These techniques are no less subject to the
relevant phase equilibria and nucleation kinetics, but they additionally
leverage transport phenomena to optimize (and in many cases circumvent)
aspects of those factors.

Throughout this Chapter, we will treat these many technologies from a
principally physical perspective, and we will adopt to a significant
degree the terminology and philosophy of Gibbsian geometric
thermodynamics. For a robust introduction to this school of thought, we
recommend the seminal textbook of Herbert Callen \cite{r4}.

\section{Thermodynamic and physical principles of ice management in
cryopreservation}
Cryopreservation, at its most fundamental, seeks to manipulate the state
and action of water in and around a biological specimen such that
mechanical, chemical, and biological damage are avoided at those
sub-normothermic temperatures at which the processes of life are slowed
or arrested. While the toxicological challenge of cryopreservation is
concerned principally with the biochemical action of this water, the
physical challenge of cryopreservation most often involves the
\emph{phase} of this water, and its tendency to change in undesirable
ways upon cooling to the sub-0 \textdegree C temperatures at which the slowing of
life processes becomes sufficient to enable days-, weeks-, or years-long
preservation. As such, our physical description of cryopreservation is
first and foremost a description of the unique behaviors of liquid water
and solid ice, in the presence of various chemicals, under various
temperatures and pressures, in contact with various surfaces, and
subject to various gradients in time, space, energy, and matter. We will
group this description into three sub-sections, treating relevant
aspects of the equilibrium thermodynamics, ice nucleation and growth
kinetics, and heat and mass transport at play.

\subsection{Equilibrium thermodynamics}

We begin our description of the thermodynamics of cryopreservation with
the most fundamental quantity dictating the response of water and ice to
these varying conditions: the chemical potential, \(\mu\). The chemical
potential, or the rate at which the free energy of a substance changes
with the addition of more of that substance to a system, is the
thermodynamic parameter most responsible for directing the action of
water. It dictates both whether water can (and in part whether it
\emph{will}) change phase from a liquid to a solid; it dictates the
extent to which (and in part \emph{the speed} with which) crystalline
ice will grow upon nucleation; and it dictates whether (and in part
\emph{how quickly}) water will migrate by diffusion or osmosis. As such,
description of the chemical potential is the first step in formulating
our broader description of physical cryopreservation.

\subsubsection{The chemical potential}

The chemical potential of a component \emph{i} is defined as:

\begin{equation}
\mu_{i} = \left( \frac{\partial G}{\partial N_{i}} \right)_{T,p,N_{j \neq i}}
\end{equation}

Wherein \emph{G} gives the Gibbs free energy of the system,
\emph{N\textsubscript{i}} is the number of mols of component \emph{i},
and subscripts outside the differentials indicate variables held
constant. \(\mu_{i}\) is most typically given in units of J/mol, and it
is an intensive thermodynamic variable, which means that it is
scale-invariant, and it provides a driving force for equilibration among
chemical species. Analogous to how differences in temperature compel the
flow of heat and differences in pressure compel the flow of volume,
differences in chemical potential compel the flow of molecules, be they
into new phases (as in the formation of ice) or into new locations (as
in osmosis or diffusion).

For pure water (or any pure substance), the chemical potential is simply
equal to the Gibbs free energy per mol. For liquid water, common
hexagonal ice (ice Ih), and several high-pressure ice polymorphs, this
quantity has been rigorously measured across a wide range of
temperatures and pressures and parameterized into powerful and
accessible equations of state. Throughout this work, we will use the
open-source SeaFreeze \cite{r5} package maintained by Journaux
and colleagues.

Thermodynamic equilibrium in a cryobiological system consisting of
different phases (or alternatively compartments) 1 and 2 requires the
condition of equality of the intensive thermodynamic variables
(\(\mu_{1} = \mu_{2}\), \(p_{1} = p_{2}\), \(T_{1} = T_{2}\)).
Practically speaking, these conditions state that, at a given pressure,
at the phase transition temperature, wherein the two phases at hand are
definitionally in equilibrium, their chemical potentials must be equal.
We refer to the set of interrelated intensive variables that define
equilibrium between two or more phases as phase equilibria, or
equilibrium coordinates.

In cryobiological problems, calculation of the chemical potential is
used for three principal purposes: 1) to solve for the temperature
(and/or pressure) at which water and ice are in equilibrium in a given
solution, i.e. to calculate the composition-dependent melting point of
ice, 2) to calculate the driving force for ice nucleation (discussed in
later sections), and 3) to calculate the driving force for mass transfer
during CPA loading (discussed in later sections).

\subsubsection{Solution theories}

As stated, in the modern era, the chemical potentials of pure water and
pure ice are well known - the challenge of cryopreservation thus falls
to describing how the chemical potential of water \emph{changes} in the
presence of the myriad biologically-necessary or cryoprotective
molecules, i.e. in solution. Armed with this knowledge, because the
chemical potential of ice does not change with solution composition
(given that ice rejects solutes and forms an approximately pure
crystal), liquid-ice phase equilibria may be readily determined by
equating the calculated chemical potential of water in solution with the
known chemical potential of ice and solving for the
temperature-pressure-concentration conditions that yield that equality.

To calculate this chemical potential, we utilize geometric or Gibbsian
solution thermodynamics. Originally derived in the late
19\textsuperscript{th} and early 20\textsuperscript{th} centuries by
Gibbs, Raoult, Lewis, Hildebrand, and others, and today practiced in
many distinct varieties, the Gibbsian school of solution thermodynamics
generally treats the free energy of a solution \emph{G} as a
mol-weighted average of the free energy of its constituent components
plus a term capturing the effects of mixing these components together,
such that

\begin{equation}
G = \sum_{i}^{}{N_{i}\mu_{i}^{0}} + \Delta G_{\mathrm{mix}}
\end{equation}

Generally, \(\Delta G_{\mathrm{mix}}\) aims to capture the effects of entropy
that accompany mixing arbitrary components together (the entropy of
mixing \(\Delta S_{\mathrm{mix}}\)) and the molecular interactions that may occur
between the components (the enthalpy of mixing \(\Delta H_{\mathrm{mix}}\)), such
that \(\Delta G_{\mathrm{mix}} = \ \Delta H_{\mathrm{mix}} - T\Delta S_{\mathrm{mix}}\). While this
framework is more or less universal within Gibbsian thermodynamics,
specific solution theories proceed to quantify \(\Delta G_{\mathrm{mix}}\) in a
variety of different ways, requiring different degrees of experimental
input and yielding different predictive and descriptive powers. We will
discuss here three key theories, chosen for their particular
applicability to problems of cryobiology and their extensibility to
multi-solute solutions.

The simplest theory of solutions is the ideal model, which assumes that
the different components in solution do not interact whatsoever
(\(\Delta H_{\mathrm{mix}} = 0\)), and that, for entropic purposes, the molecules
may be treated as physically identical. In this case, the entropy of
mixing (in units of Joules per Kelvin) is given by

\begin{equation}
\Delta S_{\mathrm{mix}}^{\mathrm{ideal}} = - R\sum_{i}^{}{N_{i}\ln x_{i}}
\end{equation}

The Gibbs free energy of the solution is then

\begin{equation}
G = \sum_{i}^{}{N_{i}\mu_{i}^{0}} + RT\sum_{i}^{}{N_{i}\ln x_{i}}
\end{equation}

and the chemical potential of component \emph{i} in an ideal solution is

\begin{equation}
\mu_{i} = \mu_{i}^{0} + RT\ln x_{i}
\end{equation}

Critically, this result implies that, within the ideal conception, the
chemical potential of a given component is compositionally dependent
only on \emph{its own} concentration, and agnostic to the specific
solutes accompanying it in solution. As cryobiologists surely
appreciate, this agnosticism does not bear out experimentally,
considering that different cryoprotectants in reality depress the
melting point of water to very differing degrees at the same molar
concentration.

The ideal model is thus limited in empirical accuracy but has
nonetheless proven an invaluable exploratory tool for the interrogation
of generalized effects of composition on various physical processes.
This utility is driven both by its simplicity and its reliance only on
properties of the \emph{pure} components, as opposed to properties of
the solution itself, which enables \emph{predictive} exploration of new
solutions without synthesizing the desired solution and measuring some
subset of its properties. Furthermore, the ideal model is readily
extensible to arbitrarily complex \emph{n}­-component solutions.

Recently, Alliston et al. \cite{r6} introduced a modification to
the ideal model (the size-dependent ideal solution or SIS model) to
incorporate solute-specific size information whilst retaining the
predictive, parameter-free nature of the ideal solution theory and its
easy extensibility to multi-component mixtures. Based on statistical
mechanics arguments by Flory \cite{r7} and Hildebrand
\cite{r8} and kinetic observations by Powell-Palm et al.
\cite{r9}, Alliston et al. replace the ideal entropy of mixing
with a size-dependent entropy of mixing capturing the entropic impact of
the considerable size difference between water and typical organic
cryoprotectants, which often possesses \textasciitilde4--10 times the
molar volume of water. For a solution of components \(i\) with molar
volumes \(v_{i}\), and volume fraction \(\phi_{i}\), the SIS mixing
entropy is

\begin{equation}
\Delta S_{\mathrm{mix}}^{\mathrm{SIS}} = - R\sum_{i}^{}{N_{i}\ln\left( \frac{{v_{i}N}_{i}}{\sum_{j}^{}{v_{j}N_{j}}} \right)} = - R\sum_{i}^{}{N_{i}\ln\phi_{i}}
\end{equation}

Substituting this entropy into the expression for the Gibbs free energy
and differentiating, one obtains a size-dependent chemical potential.
For a general multicomponent mixture, this can be written compactly as:

\begin{equation}
\frac{\mu_{i} - \mu_{i}^{0}}{RT} = \ln\phi_{i} + \sum_{j \neq i}^{}{\left( 1 - r_{\mathrm{ij}} \right)\phi_{j}}
\end{equation}

where \(r_{\mathrm{ij}} = v_{i}/v_{j}\) is the ratio of molar volumes. Using only
bulk-material molar volumes at room temperature, the incorporation of
this simple size-dependence has been shown to increase the accuracy of
the ideal model by \textasciitilde50\% or more for binary solutions of
water and cryopreservation-relevant organic molecules, both in
calculation of the melting point of ice in solution and in
identification of solubility limits and eutectic
temperatures/compositions.

While the above theories are purely \emph{predictive}, requiring no
experimental knowledge of the solution of interest, more accurate
theories can be developed using experimental input measured from a few
discrete concentrations. Many such descriptive theories start from the
standard definitions:

\begin{equation}
\mu_{i} = \mu_{i}^{0} + RT\ln a_{i}
\end{equation}

\begin{equation}
\gamma_{i} = \frac{a_{i}}{x_{i}}
\end{equation}

wherein \(a_{i}\) denotes the activity and \(\gamma_{i}\) the activity
coefficient of component \emph{i} in solution, and is a function of both
composition, temperature, and pressure (though the dependence on
temperature and pressure is often very weak). The activity coefficient
of a component is an empirical measure of its deviation from ideality
(as \(\gamma^{\mathrm{ideal}} = 1\)), and activity-based descriptive theories
typically seek to prescribe how \(\gamma_{i}\) varies with composition.

A convenient unifying language for many non-ideal models is the excess
free energy approach, which considers the free energy as the sum of ideal
and excess contributions: \(G/N = g = g^{\mathrm{ideal}} + g^{\mathrm{excess}}\). The
activity coefficient is thus given by

\begin{equation}
\ln\gamma_{i} = \frac{1}{RT}\left( \frac{\partial G^{\mathrm{excess}}}{\partial N_{i}} \right)_{T,p,N_{j \neq i}}
\end{equation}

The ideal Gibbs free energy is given in Equation 4 and the ideal entropy
is given in Equation 3. The size-dependent ideal solution theory, which
maintains zero excess enthalpy, has excess entropy
(\(Ts^{\mathrm{E}} = h^{\mathrm{E}} - g^{\mathrm{E}}\)) and water activity coefficient of

\begin{equation}
s^{\mathrm{E,SIS}} = - R\sum_{i}^{}{x_{i}\ln\left( \frac{\phi_{i}}{x_{i}} \right)}
\end{equation}

\begin{equation}
\ln\gamma_{\mathrm{w}}^{\mathrm{SIS}} = \ln\left( \frac{\phi_{\mathrm{w}}}{x_{\mathrm{w}}} \right) + \sum_{i}^{}{\left( 1 - \frac{v_{\mathrm{w}}}{v_{i}} \right)\phi_{i}}
\end{equation}

The regular solution theory represents the simplest energetic extension
of the ideal solution theory. In its conventional form, it assumes zero
excess \emph{entropy} (no size-dependence), but allows for nonzero
\emph{enthalpy} of mixing, typically represented in terms of binary
interaction parameters, \(\chi_{\mathrm{ij}}\), between each pair of species. The
excess enthalpy and water activity coefficient for a regular solution
can be written as

\begin{equation}
h^{\mathrm{E,regular}} = \frac{1}{2}RT\sum_{i,j}^{}{\chi_{\mathrm{ij}}x_{i}x_{j}}
\end{equation}

\begin{equation}
\ln\gamma_{\mathrm{w}}^{\mathrm{regular}} = \sum_{i}^{}{\chi_{\mathrm{w}i}x_{i}^{2}} + \sum_{j > i}^{}{\left( \chi_{\mathrm{w}i} + \chi_{\mathrm{w}j} + \chi_{ij} \right)x_{i}x_{j}}
\end{equation}

For binary systems this reduces to the familiar symmetric form involving
a single \(\chi\) parameter. In practice, the \(\chi_{ij}\) are fitted
to empirical data such as freezing points or vapor pressures. In the
context of cryobiology, regular solutions are rarely used in their full
generality, but the underlying idea of augmenting the ideal or SIS
baseline with simple energetic interaction terms underpins many common
approaches.

In cryobiological systems, the chemical potential of water is of primary
interest; the activities of the other solutes are needed in fewer
instances. It is therefore common to express the state of water in
solution by the parameters known as osmolality and osmotic pressure.
Osmolality is directly related to the chemical potential of water and
the water activity by

\begin{equation}
\pi = - \frac{\ln a_{\mathrm{w}}}{M_{\mathrm{w}}} = \frac{\mu_{\mathrm{w}}^{0} - \mu_{\mathrm{w}}}{M_{\mathrm{w}}RT}
\end{equation}

where \(M_{\mathrm{w}}\) is the molecular weight of water. Normal physiological
osmolality lies approximately in the range 280-300 mOsm/kg. The
corresponding osmotic pressure is related to \(\pi\) via

\begin{equation}
\Pi = \pi\rho_{\mathrm{w}}RT
\end{equation}

with \(\rho_{\mathrm{w}}\) the density of water. Normal physiological osmotic
pressure (at 37 \textdegree C) is approximately 7--8 atm.

A particularly useful implementation of regular solution theory
pioneered by Elliott and colleagues for cryobiologically-relevant
mixtures is the osmotic virial equation \cite{r10,r11,r12}. In this
model, the osmolality is expanded as a polynomial in terms of solute
molalities, \(m_{i}\):

\begin{equation}
\pi = \sum_{i}^{}{k_{i}m_{i}} + \sum_{i,j}^{}{\frac{B_{i} + B_{j}}{2}k_{i}m_{i}k_{j}m_{j}} + \sum_{i,j,k}^{}{\left( C_{i}C_{j}C_{k} \right)^{1/3}k_{i}m_{i}k_{k}m_{j}k_{k}m_{k}}
\end{equation}

Here, \(k_{i}\) is the dissociation constant (capturing ionic
dissociation of electrolytes), and \(B_{i}\) and \(C_{i}\) are the first
and second virial coefficients encapsulating pairwise and triplet
interactions between solutes. For each solute, the parameters \(k_{i}\),
\(B_{i}\), and \(C_{i}\) are obtained by fitting single-solute osmotic or
freezing-point data; multisolute behavior is then predicted via
combining rules analogous to those used in regular solution theory.
Requiring only binary parameters to describe multi-component solutions,
the osmotic virial equation has become a workhorse solution theory for
many cryobiological modeling efforts.

Beyond the models emphasized here, a wide variety of more elaborate
solution theories have been developed which could be of use to
cryobiological problems. These include classical local-composition
models such as NRTL \cite{r13} (non-random two-liquid; and
extensions such as eNRTL) and UNIQUAC\textsuperscript{13} (universal
quasi-chemical); group contribution methods built upon these such as
UNIFAC\textsuperscript{13} (UNIQUAC functional-group activity
coefficients) and AIOMFAC \cite{r14} (aerosol inorganic-organic
mixtures functional group activity coefficients); and quantum-chemical
models such as COSMO-RS \cite{r15,r16} and COSMO-SAC
\cite{r17} (conductor-like screening model). These models are
widely used in chemical engineering for vapor--liquid and liquid--liquid
equilibria and provide a semi-predictive way to handle large families of
organic solutes. As the cryobiology field moves toward rational design
of multi-component CPA cocktails (including natural deep eutectic
solvents and polymer-rich formulations), such models offer a path to
predictive screening of candidate mixtures without the need to commit to
extensive experimental testing.

\subsubsection{Phase equilibria}

\begin{figure}[h]
\centering
\includegraphics[width=1.0\textwidth]{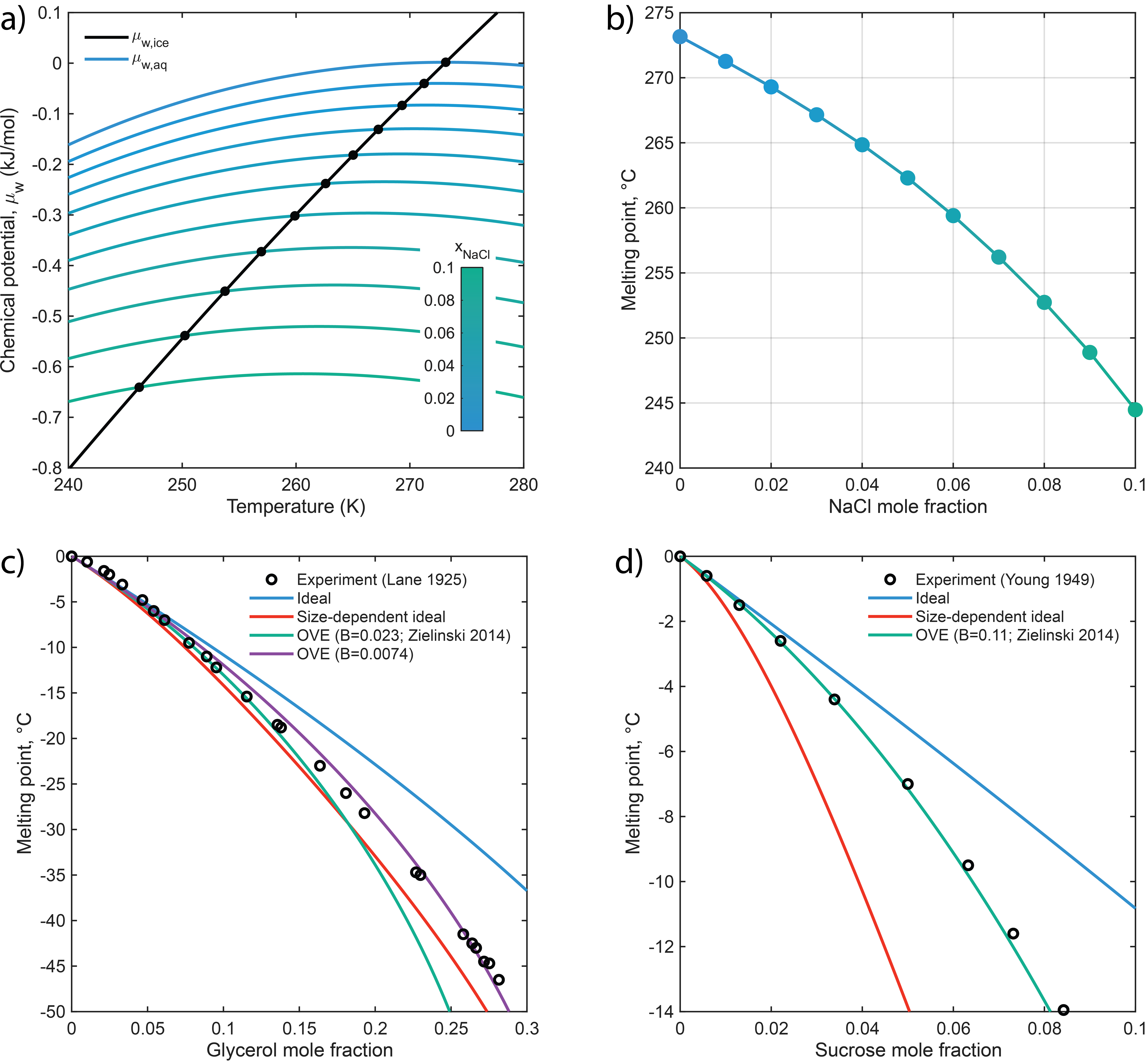}
\caption{Equilibrium melting points. a) Chemical potential of
water in solution with NaCl and in ice as a function of temperature and b) corresponding liquidus curve. Intersections of the ice and water curves denote the coexistence temperature (i.e., \(T_{\mathrm{m}}\)). Melting point as a function of mole fraction for a typical b) permeating solute (glycerol) and c) non-permeating solute (sucrose).}
\label{fig:fig1}
\end{figure}

Armed with these solution models of the chemical potential, we are now
empowered to mathematically predict or describe the melting points,
solubilities, eutectics, and other cryopreservation-relevant equilibria
of aqueous cryoprotectant solutions. We will describe two simple methods
by which to achieve this.

First, we may simply calculate the chemical potential of water in
solution at a desired concentration, using any of the theories above,
compare it to the chemical potential of ice drawn from any robust
equation of state \cite{r5}, and identify the conditions under
which the two are equal. \textbf{Fig. 1a} shows this method for
solutions of NaCl in water, using the SeaFreeze equation of state for
the chemical potential of water in ice-Ih and aqueous NaCl solutions,
with the corresponding liquidus curve shown in \textbf{Fig. 1b. Fig.
1c,d} show the predicted liquidus curves for binary solutions of
glycerol or sucrose computed using the ideal, size-dependent ideal, and
osmotic-virial equations.

A more compact route to determining the phase coexistence exploits the
Gibbs-Helmholtz relation, which for a pure substance is written as

\begin{equation}
\frac{\partial}{\partial T}\left( \frac{\mu}{RT} \right)_{\mathrm{p}} = - \frac{h}{RT^{2}}
\end{equation}

and for the difference between liquid and solid phases as

\begin{equation}
\frac{\partial}{\partial T}\left( \frac{\Delta\mu_{\mathrm{liquid - solid}}^{0}}{RT} \right)_{\mathrm{p}} = - \frac{\Delta h_{\mathrm{liquid - solid}}^{0}}{RT^{2}}
\end{equation}

where \(\Delta H_{\mathrm{liquid - solid}}^{0}\) is the enthalpy of fusion of
pure water (i.e., \(\Delta H_{\mathrm{f}}^{0}\)). At the melting point of a
solution, the liquid is in equilibrium with (practically) pure ice, so

\begin{equation}
\mu_{\mathrm{w}}^{0,\mathrm{ice}}\left( T_{\mathrm{m}} \right) = \mu_{\mathrm{w,aq}}\left( T_{\mathrm{m}},a_{\mathrm{w}} \right) = \mu_{\mathrm{w}}^{0,\mathrm{liq}}\left( T_{\mathrm{m}} \right) - \pi RT_{\mathrm{m}}M_{\mathrm{w}}
\end{equation}

Hence

\begin{equation}
\pi = \frac{1}{M_{\mathrm{w}}R}\int_{T_{\mathrm{m}}}^{T_{\mathrm{m}}^{0}}{\frac{\Delta H_{\mathrm{f}}^{0}}{T^{2}}dT}
\end{equation}

At the melting point, the enthalpy and entropy are related by
\(\Delta H_{\mathrm{f}}^{0} = T\Delta S_{\mathrm{f}}^{0}\), and so the entropy of fusion
can be substituted in the place of enthalpy in these relations. This
integral form of the above expression is exact for a given
\(\Delta H_{\mathrm{f}}^{0}(T)\), but the temperature dependence of enthalpies
may not always be known. To a first-order approximation, the enthalpy of
fusion may be assumed independent of temperature
(\(\Delta H_{\mathrm{f}}^{0}(T) \approx \Delta H_{\mathrm{f}}^{0}(T_{\mathrm{m}})\)) yielding the
0\textsuperscript{th}-order Gibbs-Helmholtz liquidus relation:

\begin{equation}
\pi \approx \frac{\Delta H_{\mathrm{f}}^{0}}{M_{\mathrm{w}}R}\left( \frac{1}{T_{\mathrm{m}}} - \frac{1}{T_{\mathrm{m}}^{0}} \right)
\end{equation}

This may also be rearranged to provide a closed-form expression for the
dependence of the coexistence temperature on the osmolality (or solvent
activity in the more general case):

\begin{equation}
T_{\mathrm{m}} \approx \left( \frac{1}{T_{\mathrm{m}}^{0}} + \frac{\pi RM_{\mathrm{w}}}{\Delta H_{\mathrm{f}}^{0}} \right)^{- 1}
\end{equation}

Due to the wide availability of tabulated enthalpy of fusion values, the
0\textsuperscript{th}-order Gibbs-Helmholtz relation is particularly
useful. A more accurate (though data-intensive) description involves a
first-order expansion of the enthalpy of fusion via the heat capacity
(\(\Delta H_{\mathrm{f}}(T) \approx \Delta H_{\mathrm{f}}^{0} + \left( T - T_{\mathrm{m}}^{0} \right)\Delta C_{\mathrm{p}}^{0}\)),
yielding

\begin{equation}
\pi \approx \frac{\Delta H_{\mathrm{f}}^{0}}{M_{\mathrm{w}}R}\left( \frac{1}{T_{\mathrm{m}}} - \frac{1}{T_{\mathrm{m}}^{0}} \right) + \frac{\Delta C_{p,\mathrm{f}}^{0}}{M_{\mathrm{w}}}\left( \frac{T_{\mathrm{m}}^{0}}{T_{\mathrm{m}}} - \ln\left( \frac{T_{\mathrm{m}}^{0}}{T_{\mathrm{m}}} \right) - 1 \right)
\end{equation}

From Equation 22 we can also derive the cryoscopic constant for water,
\(K_{\mathrm{f}}\) = --1.86 K per osmole/kg, which is a proportionality constant
relating the melting point depression (relative to pure water) to the
solution osmolality:

\begin{equation}
K_{\mathrm{f}} = \frac{R{T_{\mathrm{m},0}}^{2}M_{\mathrm{w}}}{\Delta H_{\mathrm{f}}^{0}} \approx \frac{\Delta T_{\mathrm{m}}}{\pi}
\end{equation}

With these relations it is readily possible to compute the melting point
of a solution given the osmolality (or the osmolality given the melting
point). The 0\textsuperscript{th}-order approximation remains accurate
to within 1 \textdegree C down to --28 \textdegree C and both the first-order approximation
and cryoscopic ratio down to roughly --52 \textdegree C.

One exemplary application of the first-order Gibbs-Helmholtz relation is
the computation of eutectic phase diagrams (liquidus and solidus curves
inclusive) using only pure component data (\(\Delta H_{\mathrm{f}}^{0}\) and
\(T_{\mathrm{m}}^{0}\)). Alliston, et al.\textsuperscript{6} recently reported
that the size-dependent 0\textsuperscript{th}-order Gibbs-Helmholtz
relation provides remarkably accurate predictions of binary eutectic
phase transitions in cryobiologically relevant binary solutions.
Examples for glycerol and sucrose are depicted in \textbf{Fig. 2}.

\begin{figure}[t]
\centering
\includegraphics[width=1.0\textwidth]{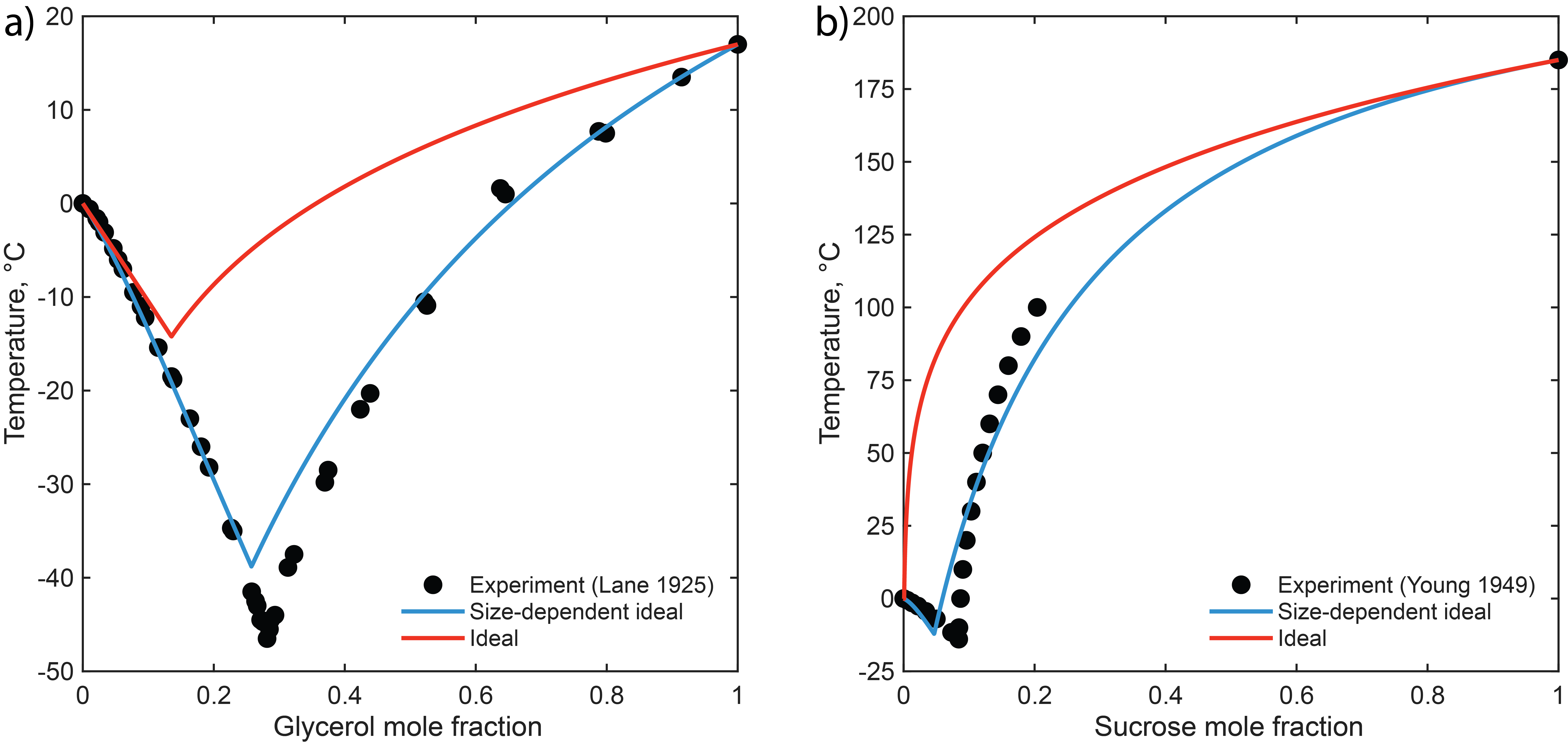}
\caption{Eutectic phase diagrams for binary aqueous solutions
generated via size-dependent and ideal predictive models. Experimental data from the literature is shown in gray markers for a) glycerol and b) sucrose. Both models require only pure component properties, and do not require empirical information on the solution itself.}
\label{fig:fig2}
\end{figure}

\subsubsection{Isochoric phase equilibrium}

Thus far much of the discussion has considered systems under isobaric
conditions---those in which the system is in contact with a pressure
reservoir such as the atmosphere, and for which volume is able to freely
vary. In isochoric (constant volume or confined) systems \cite{r18,r19}, the \emph{volume} of the system is controlled instead, and the
\emph{pressure} is free to vary. This thermodynamic situation arises
when a sample is cryopreserved within a rigid, hermetically sealed
container absent air (or any bulk gas phase), referred to in recent
literature as an isochoric chamber. Under isochoric conditions, the
natural thermodynamic variables are \(\{ T,V,x\}\), and the according
thermodynamic potential describing the system is the Helmholtz free
energy. Thus, the equilibrium conditions become

\begin{equation}
\mu_{\mathrm{w,aq}} = \mu_{\mathrm{w,ice}}\ :\ \ \ \ \mu_{\mathrm{w,aq}}^{\mathrm{v}} - {\overline{v}}_{\mathrm{w,aq}}\left( \frac{\partial\mu_{\mathrm{w,aq}}^{\mathrm{v}}}{\partial{\overline{v}}_{\mathrm{w,aq}}} \right)_{T,x} = \mu_{\mathrm{w,ice}}^{\mathrm{v}} - {\overline{v}}_{\mathrm{w,ice}}\left( \frac{\partial\mu_{\mathrm{w,ice}}^{\mathrm{v}}}{\partial{\overline{v}}_{\mathrm{w,ice}}} \right)_{T,x}
\end{equation}

\begin{equation}
P_{\mathrm{aq}} = P_{\mathrm{ice}}:\ \ \ \ \left( \frac{\partial\mu_{\mathrm{w,aq}}^{\mathrm{v}}}{\partial{\overline{v}}_{\mathrm{w,aq}}} \right)_{T,x} = \left( \frac{\partial\mu_{\mathrm{w,ice}}^{\mathrm{v}}}{\partial{\overline{v}}_{\mathrm{w,ice}}} \right)_{T,x}
\end{equation}

\begin{equation}
T_{\mathrm{aq}} = T_{\mathrm{ice}}
\end{equation}

where \(\overline{v}\) is the partial molar volume and \(\mu_{\mathrm{w}}^{\mathrm{v}}\)
is the partial molar Helmholtz free energy. These equalities are
supplemented by the condition that the sum of the volumes of individual
phases must equal the system volume (i.e.,
\(v_{\mathrm{sys}} = (1 - f_{\mathrm{ice}})v_{\mathrm{aq}} + f_{\mathrm{ice}}v_{\mathrm{ice}}\), with \(f_{\mathrm{ice}}\) as
the volumetric phase fraction of ice).

\textbf{Fig. 3a} shows a schematic convex-hull construction
\cite{r20} of the Helmholtz free energy curves for several phases
of water under isochoric conditions, and \textbf{Fig. 3b} shows the
resulting isochoric \(T\)--\(v\) phase diagram for pure water. For many
specific volumes, cooling at fixed \(v\) necessarily drives the system
into a two-phase (ice + liquid) coexistence region, even when the
corresponding isobaric path would remain single-phase. Practically, this
enables preservation of biological matter at sub-0 \textdegree C temperatures in
the portion of the isochoric system that remains liquid, thereby
protecting it from ice formation. For a thorough review of both the
theory and practice of isochoric cryopreservation, we refer the reader
to Consiglio et al. \cite{r19}.

\begin{figure}[t]
\centering
\includegraphics[width=1.0\textwidth]{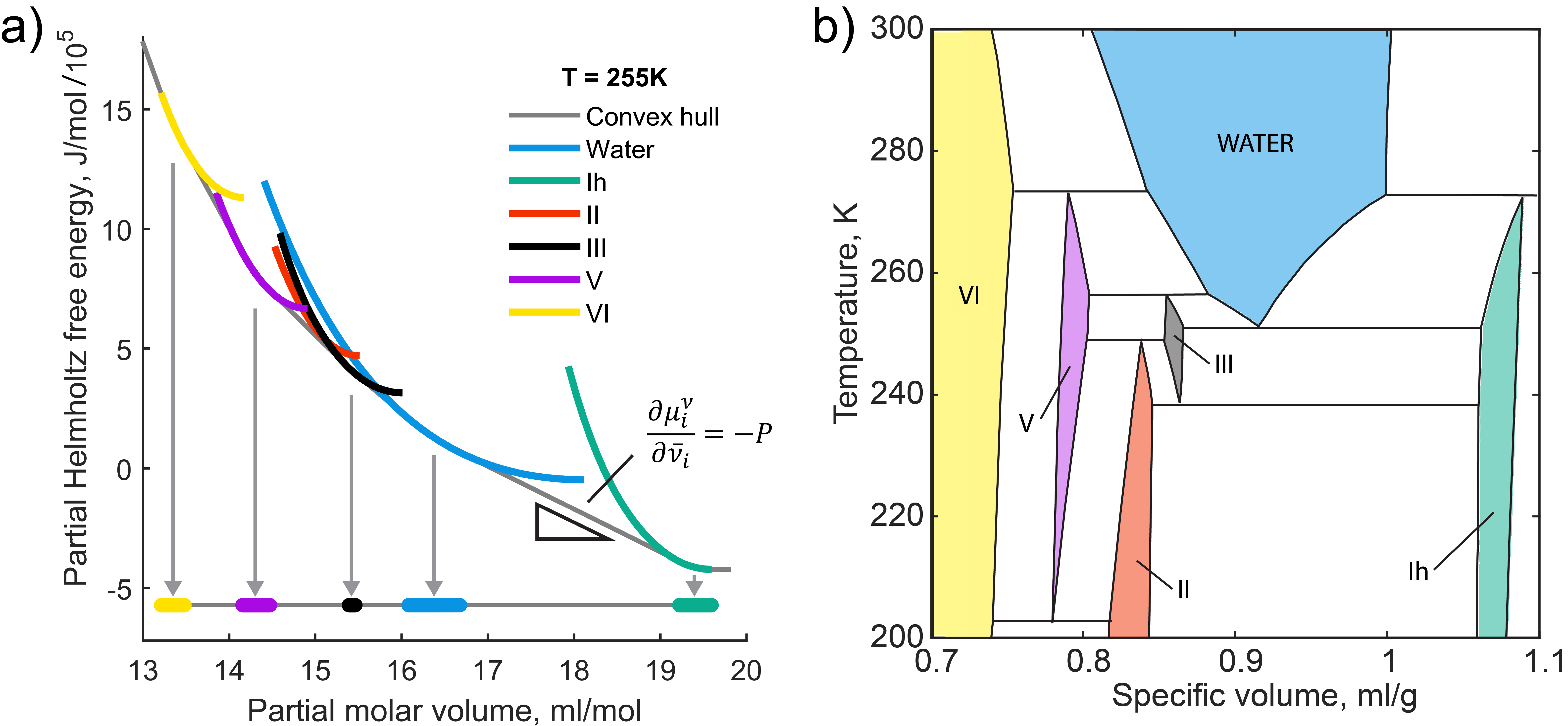}
\caption{Isochoric equilibrium: a) Convex hull construction for
determination of phase equilibria. b) Isochoric T-V phase diagram for
pure water. Adapted with permission from Consiglio et al.
\cite{r19}.}
\label{fig:fig3}
\end{figure}

\subsubsection{Phase fraction and the lever rule}

In addition to the temperature at which ice may begin to form,
cryobiologists are also often interested in the \emph{amount} of ice
that will form at a given temperature. For a binary solution at constant
pressure, a typical \(T\)--\(x\) phase diagram produces a two-phase
liquid-ice region beneath the liquidus curve. A system with overall
solute mole fraction \(x_{0}\) cooled to a temperature \(T\) such that
the coordinates {[}\(T,x_{0}\rbrack\) are inside this two-phase region
(\textbf{Fig. 4a}) will separate into distinct phases with liquid and
solid solute concentrations \(x_{\mathrm{liquid}}\) and \(x_{\mathrm{solid}}\).
Conservation of mass gives:

\begin{equation}
x_{0} = f_{\mathrm{liquid}}x_{\mathrm{liquid}} + \left( 1 - f_{\mathrm{liquid}} \right)x_{\mathrm{solid}}
\end{equation}

where \(f_{\mathrm{liquid}}\) is the phase fraction that remains liquid (in units
that match the concentration, be they weight fraction, volume fraction,
mol fraction, etc.). In aqueous solutions, the solute concentration in
the solid ice phase is typically taken as \(x_{\mathrm{solid}} \approx 0\),
because ice rejects nearly all solutes. As such, the unfrozen phase
fraction \(f_{\mathrm{liquid}}\) for aqueous solutions simplifies to

\begin{equation}
f_{\mathrm{liquid}} = \frac{x_{0}}{x_{\mathrm{liq}}}
\end{equation}

As depicted in \textbf{Fig. 4a}, the value of \(x_{\mathrm{liq}}\) is given by
the concentration coordinate of the liquidus curve at the sub-freezing
temperature \(T\), and this same expression may be derived by graphical
examination of the phase diagram, using what is commonly referred to as
the ``lever rule''.

In \textbf{Fig. 4b} we show the unfrozen (liquid) phase fraction
\(f_{\mathrm{liquid}}\) versus temperature for water-DMSO solutions at various
starting DMSO concentrations \(x_{0}\). For small starting \(x_{0}\), a
significant fraction of the system quickly transforms to ice as the
temperature falls below the liquidus curve, but the rate-of-change of
this fraction with temperature decays continuously as deeper
temperatures are reached. This behavior is explained by, \textbf{Fig.
4c}, which shows the corresponding liquid-phase concentration
\(x_{\mathrm{liquid}}(T)\), and illustrates how solute ``ripening'' from
rejection by the solid phase intensifies as freezing progresses. In
classical slow-freezing cryopreservation protocols, this progressive
concentration of the remaining liquid is deliberately exploited: as
extracellular ice forms and solutes are rejected, the external solution
becomes more concentrated, drawing water osmotically out of cells and
thereby reducing intracellular ice formation. When carried too far,
however, the same process leads to extreme extracellular solute
concentrations and osmotic stresses, producing the ``solution effects''
injury central to Mazur's two-factor hypothesis of freezing damage
\cite{r21}.

In an isochoric system filled with pure water, the same logic applies,
but the conserved quantity is specific volume rather than composition.
For a given temperature \(T\) and total system specific volume
\(v_{\mathrm{sys}}\) lying inside the two-phase dome of the isochoric
\(T\)--\(v\) diagram, the coexisting liquid and solid phases have
specific volumes \(v_{\mathrm{liquid}}\ \)and \(v_{\mathrm{solid}}\). Conservation of
volume yields

\begin{equation}
v_{\mathrm{sys}} = f_{\mathrm{liquid}}v_{\mathrm{liquid}} + \left( 1 - f_{\mathrm{liquid}} \right)v_{\mathrm{solid}}
\end{equation}

So that

\begin{equation}
f_{\mathrm{liquid}} = \frac{v_{\mathrm{solid}} - v_{\mathrm{sys}}}{v_{\mathrm{solid}} - v_{\mathrm{w}}}
\end{equation}

\textbf{Fig. 5} (isochoric lever rule) illustrates this construction
graphically: \textbf{Fig. 5}a shows the isochoric lever rule on a
specific-volume diagram, and \textbf{Fig. 5b} shows the resulting liquid
fraction as a function of temperature for several choices of total
specific volume. In practice, the specific volume of an isochoric system
is set when the chamber is sealed---typically at 5--10 \textdegree C, giving
\(v_{0} \approx 1\text{\ mL/g}\). Other specific volumes can be accessed
by deliberately including a vapor space (increasing \(v_{0}\)) or by
compressing the system with a piston (decreasing \(v_{0}\)). If the
liquid filling the isochoric chamber is not pure water, but instead an
aqueous solution, a 3-dimensional lever rule may be devised which
conserves both phase volume \emph{and} phase concentration.

\begin{figure}[t]
\centering
\includegraphics[width=1.0\textwidth]{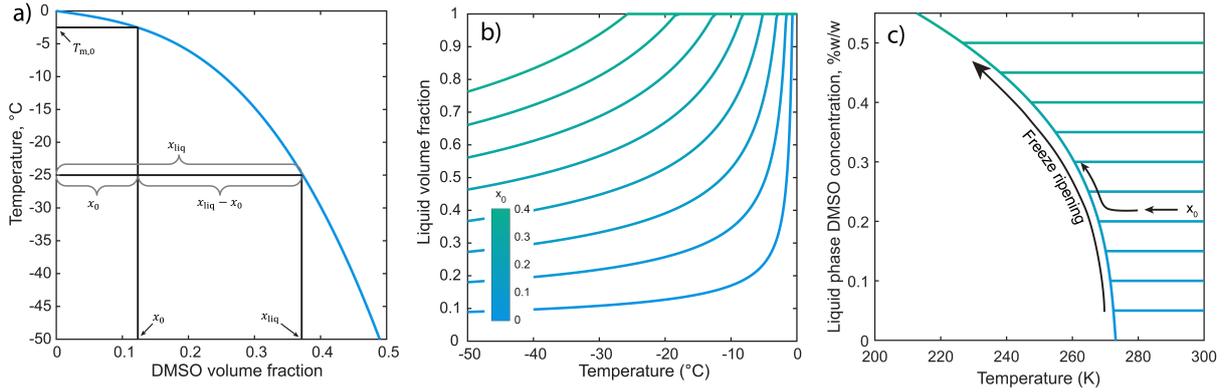}
\caption{Isobaric Ice-liquid phase fractions and solute ripening
processes. a) Construction of the lever rule for a binary solution T-x phase diagram, which enables calculation of the fraction of the system that remains in the frozen or unfrozen state at a given temperature. b) Unfrozen phase fraction versus temperature for various starting concentrations of DMSO in a water-DMSO solution. c) According concentration of the liquid phase during progressive freezing, for the same starting concentrations indicated by the color bar in panel (b).}
\label{fig:fig4}
\end{figure}

In summary, the tools of multi-phase equilibrium thermodynamics---regardless of the specific theories used to compute chemical potential---allow us to map solution composition (and, in isochoric systems,
specific volume) to melting points, eutectics, and equilibrium phase
fractions. These maps in turn underpin the thermodynamic design and
analysis of equilibrium-dominated cryopreservation strategies such as
liquidus tracking, partial freezing, and isochoric freezing, and they
set the backdrop against which kinetic and transport phenomena act in
more complex protocols.

\subsection{Ice nucleation and growth}

Having established how equilibrium thermodynamics dictates the
conditions under which liquid water and ice can coexist, we next
consider how a system \emph{transitions} between these phases. In
practice, freezing proceeds by the stochastic formation of microscopic
ice nuclei, followed by their subsequent growth. The standard framework
for describing this process is classical nucleation theory (CNT), which
links the statistics of ice nucleation directly to the underlying
solution thermodynamics and interfacial properties. We will here review
salient aspects of this theory, emphasizing both its practical
applications in cryopreservation and its connection to equilibrium
thermodynamics.

\begin{figure}[t]
\centering
\includegraphics[width=1.0\textwidth]{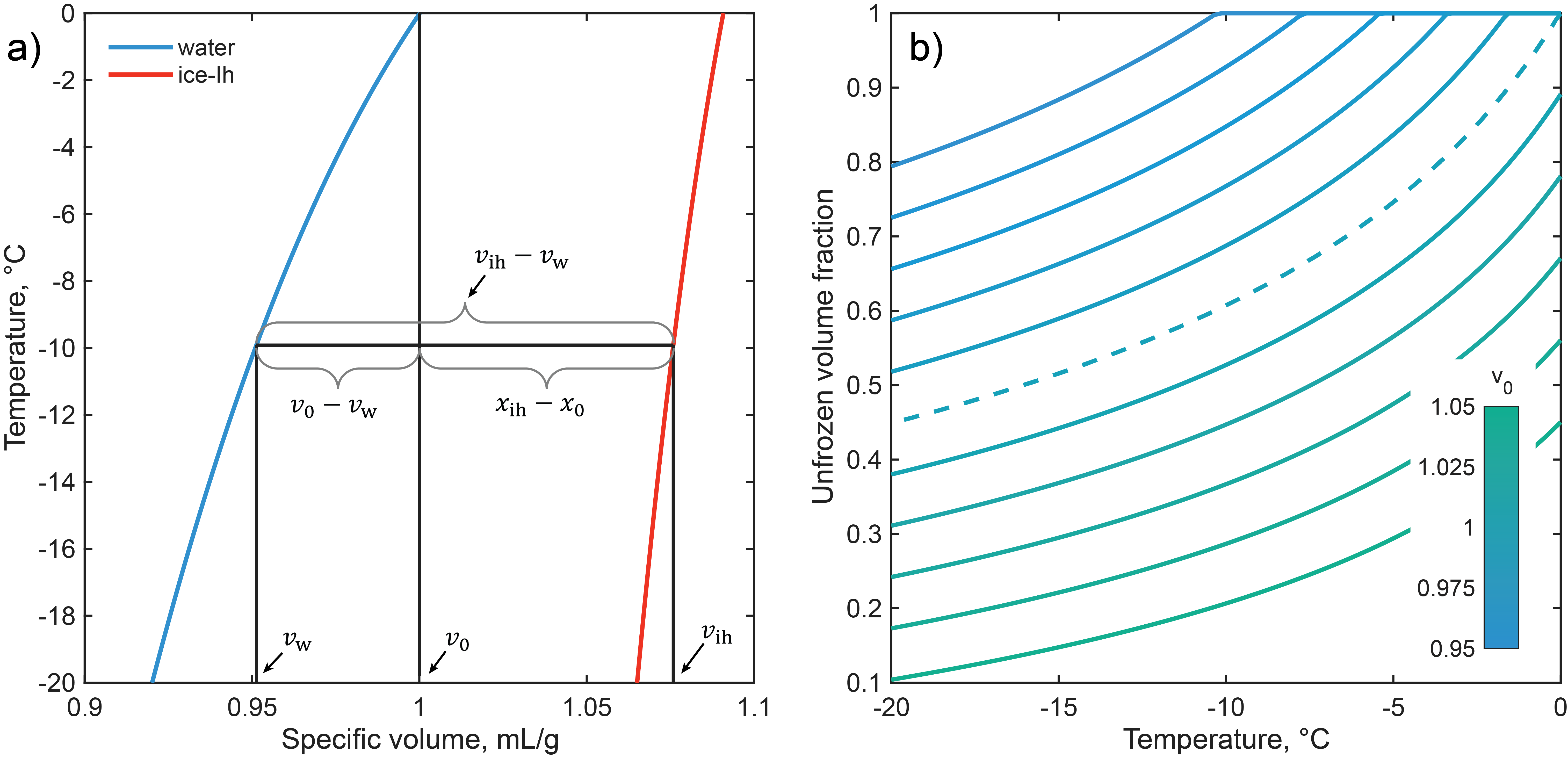}
\caption{Isochoric ice-liquid phase fractions for pure water. a)
Construction of the isochoric lever rule from the water/ice Ih
temperature-specific volume phase diagram derived by Powell-Palm et al. \cite{r20}. b) Fraction of isochoric system that remains liquid
during isochoric freezing as a function of system specific volume.}
\label{fig:fig5}
\end{figure}

\subsubsection{The nucleation barrier}

The formative relations of CNT are derived by considering the change in
energy in a system between a first state in which the entire system is
liquid and a second state in which a nucleus (a stable cluster of
molecules) of a solid phase (here ice) has emerged (\textbf{Fig. 6}).

Assuming the volume of the system is much larger than that occupied by
the emerging solid, the free energy of these two states may then be
written as:

\begin{equation}
G_{1} = \mu_{\mathrm{w,aq}}^{1}N_{\mathrm{w,aq}}^{1} + \mu_{\mathrm{s,aq}}^{1}N_{\mathrm{s,aq}}^{1}
\end{equation}

\begin{equation}
G_{2} = \mu_{\mathrm{w,aq}}^{2}N_{\mathrm{w,aq}}^{2} + \mu_{\mathrm{s,aq}}^{2}N_{\mathrm{s,aq}}^{2} + \mu_{\mathrm{w,ice}}^{2}N_{\mathrm{w,ice}}^{2} + 4\pi R^{2}\sigma_{\mathrm{iw}}
\end{equation}

\begin{figure}[h]
\centering
\includegraphics[width=0.9\textwidth]{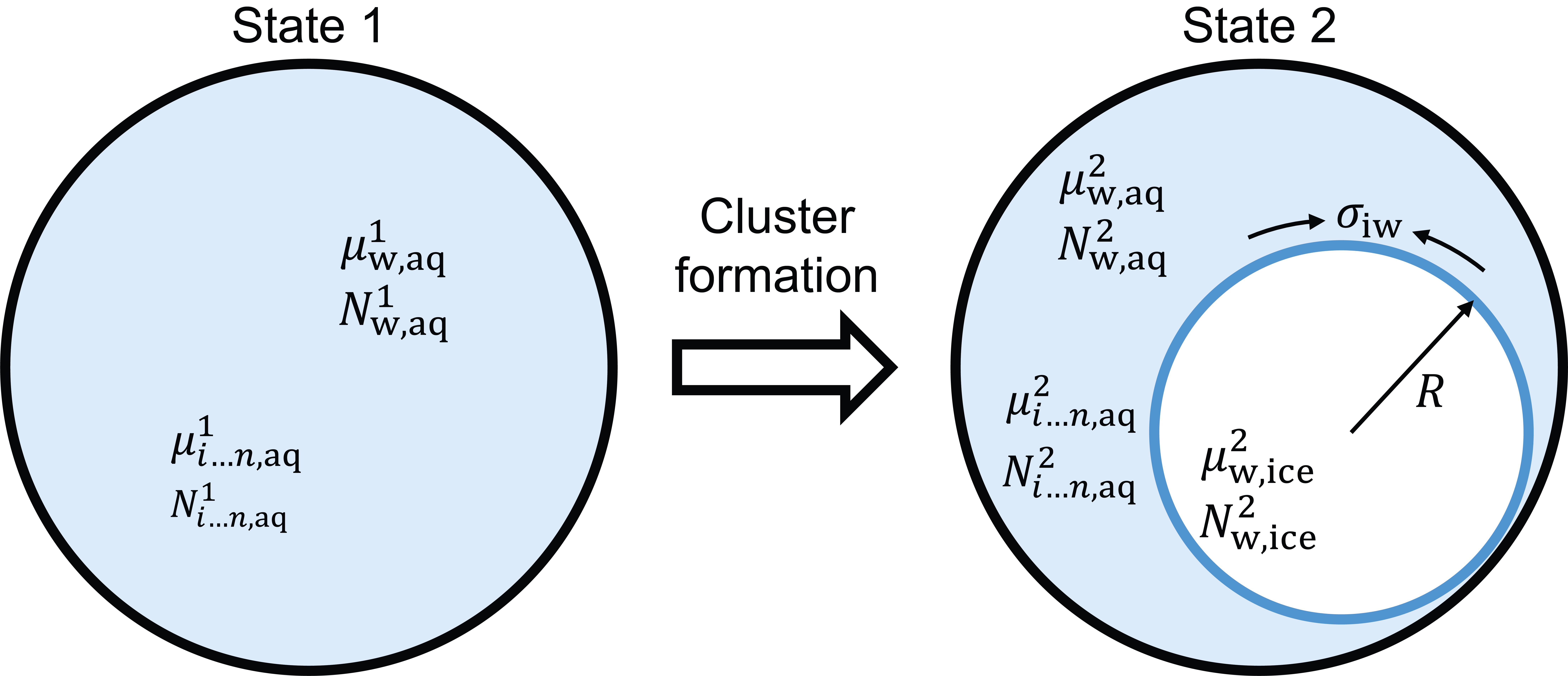}
\caption{Conceptual illustration of the change in free energy
upon formation of a solid nucleus from a supercooled liquid.}
\label{fig:fig6}
\end{figure}

In State 1, \(\mu_{\mathrm{w,aq}}^{1}N_{\mathrm{w,aq}}^{1}\) gives the free energy of bulk
liquid water in solution, and \(\mu_{\mathrm{s,aq}}^{1}N_{\mathrm{s,aq}}^{1}\) gives that
of the solute. In State 2, the free energy of the bulk incipient ice
phase is further given by \(\mu_{\mathrm{w,ice}}^{2}N_{\mathrm{w,ice}}^{2}\).

The last term in \(G_{2}\), which encodes the most critical
phenomenological insight provided by CNT, gives the \emph{surface
energy} of the ice nucleus, or the energetic and entropic toll exacted
upon the system in order to host two phases where previously there was
one. The proportionality constant of this energy is given by the
\emph{liquid-ice interfacial free energy} \(\sigma_{\mathrm{iw}}\), which
describes the energy required per unit surface area of the ice nucleus
to bound it from the liquid phase.

Again, assuming a small initial ice nucleus relative to the volume of
the system, we may take \(\mu_{\mathrm{w,aq}}^{1} \approx \mu_{\mathrm{w,aq}}^{2}\) and
\(\mu_{\mathrm{s,aq}}^{1} \approx \mu_{\mathrm{s,aq}}^{2}\). It is customary (if arbitrary) also to treat the emerging ice nucleus as spherical, with
volume \(V = \frac{4}{3}\pi R^{3} = v_{\mathrm{ice}}N\) and surface area
\(SA = 4\pi R^{2} = (36\pi)^{1/3}V^{2/3}\). Armed with these assumptions, we may now write the \emph{difference} in free energy upon spontaneous formation of an ice nucleus (the difference between States 1 and 2) as:

\begin{equation}
\Delta G_{\mathrm{nuc}} = G_{2} - G_{1} = \frac{4}{3}\pi R^{3}\left( \frac{\Delta\mu}{v_{\mathrm{ice}}} \right)\  + 4\pi R^{2}\sigma_{\mathrm{iw}}
\end{equation}

Critically, the two terms comprising this equation exist in competition.
Given the lower free energy of solid ice relative to liquid water at any
sub-melting temperature, the first term (often referred to as the
``bulk'' term) is negative, and scales with the \emph{volume} of the
growing ice phase, representing the spontaneous drive of the system to
convert to the more stable phase. Meanwhile, because the segregation of
an ordered crystalline phase from the disordered liquid must necessarily
\emph{reduce} the entropy of the system, the second term (often referred
to as the ``surface'' term) must accordingly be positive,
\emph{increasing} the free energy of the system and scaling with the
\emph{surface area} of the nucleus.

It is this surface-versus-volume competition that belies the nature of
nucleation as an activated stochastic process, and gives the nucleation
free energy difference \(\Delta G_{\mathrm{nuc}}\) its characteristic maximum
(\textbf{Fig. 7}), referred to as the nucleation barrier
\(\Delta G_{\mathrm{nuc}}^{*}\):

\begin{equation}
\Delta G_{\mathrm{nuc}}^{*} = \left. \ \Delta G_{\mathrm{nuc}} \right|_{\frac{\partial\Delta G_{\mathrm{nuc}}}{\partial N} = 0} = \frac{16\pi\sigma_{\mathrm{iw}}^{3}v_{\mathrm{ice}}^{2}}{3\Delta\mu^{2}}
\end{equation}

The random thermal motion of water molecules in the solution causes the
system to fluctuate and sample a variety of possible configurations,
which may include clusters of water molecules arranged in the cubic or
tetrahedral structures defining ice Ic or Ih. The cluster size
corresponding to \(\Delta G_{\mathrm{nuc}}^{*}\) gives ``critical'' cluster size
(or ``critical radius'' if not discretizing by number of molecules), and
the fate of any spontaneously assembled cluster of molecules is
prescribed by its relation to this critical size.

Any smaller (``sub-critical'') cluster, once assembled by a random
fluctuation, in order to then decrease its free energy and march towards
equilibrium, will be compelled to dissolve, as recruitment of additional
molecules would \emph{increase} its free energy in defiance of the
2\textsuperscript{nd} Law of Thermodynamics. Equally, any cluster
produced by random fluctuation that is \emph{larger} than the critical
size will be compelled to \emph{grow} continuously until equilibrium is
reached, initiating bulk phase crystallization or freezing. It is this
critical or supra-critical cluster that we refer to as a stable ice
\emph{nucleus.}

\begin{figure}[h]
\centering
\includegraphics[width=0.67\textwidth]{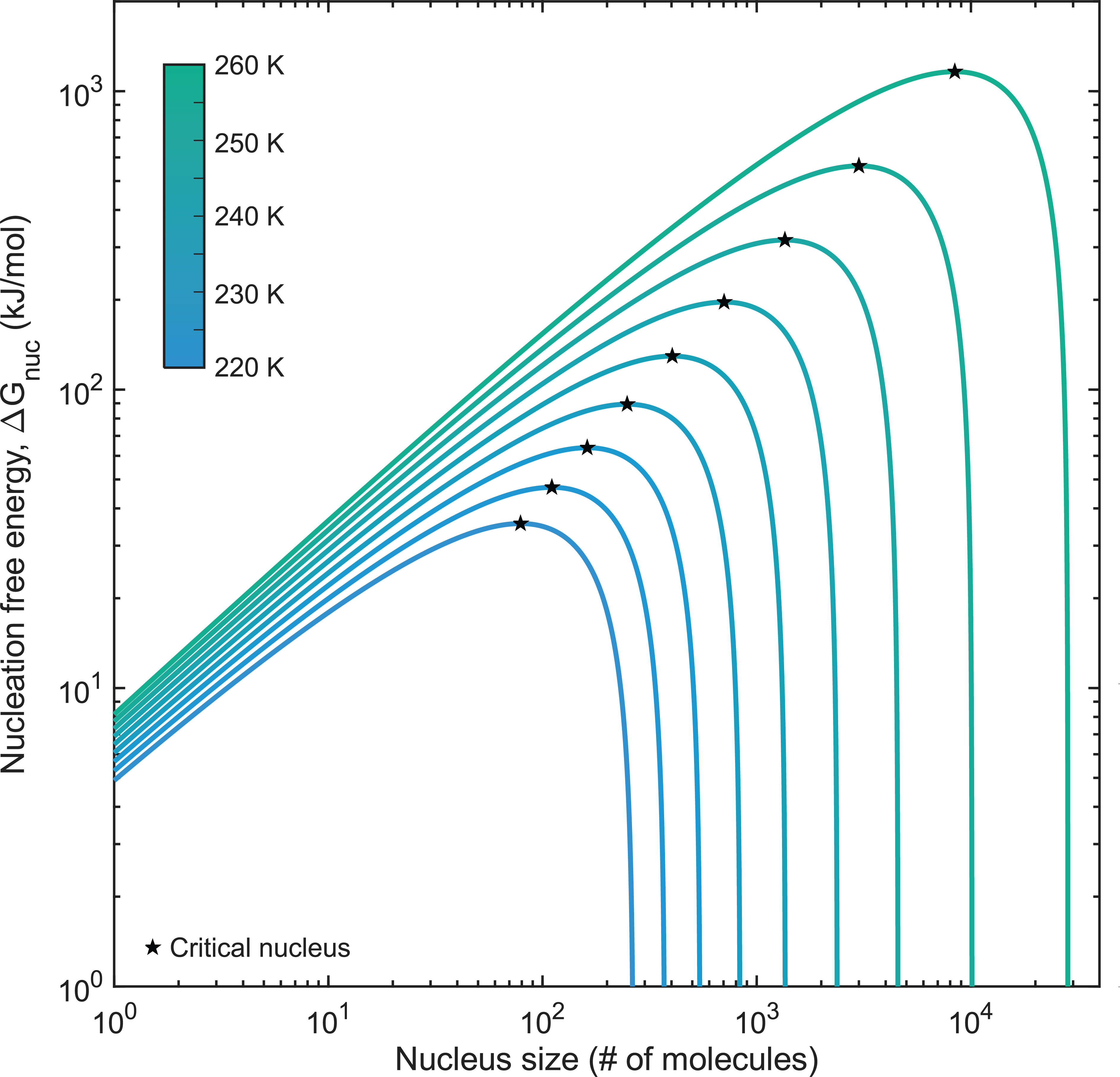}
\caption{Free energy change upon nucleation versus ice cluster
size for pure water at various temperatures. The nucleation barrier for each temperature is marked by a black star. Consistent with expectation, the nucleation barrier grows exponentially higher as the equilibrium melting temperature is approached, at which point it reaches singularity due to the chemical potential difference in the denominator of Equation
36 reaching zero.}
\label{fig:fig7}
\end{figure}

\subsubsection{The nucleation rate}

The nucleation barrier \(\Delta G_{\mathrm{nuc}}^{*}\) connects the process of
ice nucleation to the underlying bulk thermodynamics of the aqueous
solution, via the chemical potential difference \(\Delta\mu\) between
solid ice and liquid water in solution, and to the surface
thermodynamics of the solution via \(\sigma_{\mathrm{iw}}\). In isolation, this
barrier provides only relative or trend-level insight into the ease or
difficulty of ice nucleation in a given solution at a given temperature.
To valorize this information, we may then incorporate this barrier as a
rate constant in a probabilistic description of a stochastic activated
process (i.e. one driven by random molecular fluctuation), yielding a
\emph{nucleation rate J} {[}nuclei formed per m\textsuperscript{3} per
second{]}:

\begin{equation}
J = J_{0}\exp\left( - \frac{\Delta G_{\mathrm{nuc}}^{*}}{k_{\mathrm{B}}T} \right)
\end{equation}

The nucleation rate \(J\) is comprised of two general terms, the
exponential driver of the process
\(\exp\left( - \frac{\Delta G_{\mathrm{nuc}}^{*}}{k_{\mathrm{B}}T} \right)\) and the
pre-factor \(J_{0} = \frac{Zf^{*}}{v_{0}}\). We refer to the exponential
portion as the thermodynamic term, describing a simple classical
Arrhenius-style kinetic process for which the characteristic constant is
the nucleation barrier. This term represents the probability of a token
stochastic fluctuation achieving the critical cluster size. We refer to
the pre-factor \(J_{0}\) as the kinetic term, which in turn captures the
aspects of the nucleation process that are not captured by connections
to bulk thermodynamics, but instead by microscale transport processes.
Here, \(f^{*}\) gives the attachment frequency factor, which describes
the probability of a given cluster recruiting additional water molecules
based on barriers to diffusion in the solution, the Zeldovich factor
\(Z\) gives the likelihood that a nucleus of precisely the critical size
will either dissolve or grow (``roll'' to the left or right of the
maximum of the nucleation free energy curve in \textbf{Fig. 7}), and
\(v_{0}\) is the molecular volume of ice.

The nucleation rate may thus be thought of as a constant competition
between the kinetic and diffusive terms. At higher temperatures, where
(for cryobiological applications) solution viscosity is typically
relatively low (and thus diffusivity relatively high), the nucleation
process is dominated by the kinetic term, and therefore the underlying
solution thermodynamics. As the temperature decreases towards the glass
transition and the viscosity increases, however, the diffusive term will
eventually come to dominate, preventing the recruitment of new molecules
to a potentially critical nucleus, and thereby hindering the nucleation
process. This transition of dominance between the two terms is embodied
in temperature-time-transformation curves, which show that for any
material, there exists a maximum in nucleation rate located at some
temperature between that of melting and that of glass transition
\cite{r22}.

The rate equation \emph{J} further reveals the critical role of the
surface effects in the nucleation process; \emph{J} is an exponential
function of the water-ice interfacial free energy \(\sigma_{\mathrm{iw}}\) to the
third power, introducing extraordinary sensitivity to this parameter.
For example, assuming typical values of \emph{J\textsubscript{0}} for
pure water \cite{r23}, a 10\% change in the value of
\(\sigma_{\mathrm{iw}}\) can change the nucleation rate by \textgreater10 orders
of magnitude. This sensitivity renders accurate prediction of the
precise nucleation rate difficult, and recent studies have suggested
that targeting this parameter via molecular or solution design
strategies may present a potent engineering lever with which to control
cryobiological ice nucleation \cite{r9}.

\subsubsection{Growth after nucleation}

After a critical ice nucleus forms, bulk crystallization will proceed.
For pure water, the growth rate of ice in this phase is often modeled
using the Wilson-Frenkel relation \cite{r24}:

\begin{equation}
U(T) = \frac{D}{a}\left\lbrack 1 - \exp\left( - \frac{|\Delta\mu|}{RT}\  \right) \right\rbrack
\end{equation}

Wherein \(D\) is the water diffusivity, \(\Delta\mu\) is the chemical
potential difference between water in the liquid phase and ice, and
\(a\) is characteristic length often specified as the interfacial
thickness or water molecular mean free path length. Like the nucleation
rate, at small supercoolings, \(U\) is limited by the thermodynamic
driving force (via \(\Delta\mu\)), whereas at deeper supercoolings it is
limited by diffusion (via \(D\)). For cellular systems, the interplay
between growth kinetics and cellular water transport rate controls
whether ice remains extracellular or propagates into the intracellular
space \cite{r25}.

For aqueous solutions, the same basic Wilson--Frenkel concept can be
applied if we replace the pure-water driving force with the
chemical-potential difference of water between the solid and the local
solution at the interface. In local equilibrium models,
\(\mu_{\mathrm{w,ice}}(T, p)\) is set equal to \(\mu_{\mathrm{w,liq}}(T, p, x_{i})\)
along the interface, where \(x_{i}\) is the interfacial solute mole
fraction. As ice advances and rejects solute into the liquid, \(x_{i}\)
departs from the bulk composition \(x_{\infty}\), and a solute-enriched
boundary layer develops in front of the interface. The resulting local
freezing-point depression reduces \(\Delta\mu\) and progressively slows
growth, even at fixed ambient temperature.

Mathematically, sharp-interface models treat growth in solutions by
coupling the interface-kinetic relation to conservation laws for heat
and solute transport. In a one-dimensional binary solution, the solute
concentration \(c(z,\ t)\) obeys a diffusion equation ahead of a moving
interface \(z\  = \ s(t)\), together with a Stefan condition
\(\left. \ D\frac{\partial c}{\partial z} \right|_{z = s^{+}} = \ V_{\mathrm{n}}\ (c_{\mathrm{l}}\  - \ c_{\mathrm{s}})\)
that enforces solute balance at the front (here \(V_{\mathrm{n}}\) is the
interface-normal velocity and \(c_{\mathrm{l}}\) and \(c_{\mathrm{s}}\) are the solute
concentrations in the liquid and solid). When the solid phase excludes
solute (typically \(c_{\mathrm{s}} \approx 0\) for ice), this boundary condition
ensures that growth is accompanied by a concentrated liquid layer whose
thickness and composition are controlled by the competition between
interface motion (\(V_{\mathrm{n}}\)) and diffusion (\(D\)). In the
diffusion-limited regime \(V_{\mathrm{n}}\) is set by how rapidly this boundary
layer can be relaxed, whereas in the interface-limited regime \(V_{\mathrm{n}}\)
is controlled primarily by attachment kinetics at the ice surface.

A convenient way to summarize these regimes is to view the observed
growth rate as the harmonic mean of an interface-limited velocity
\(V_{\mathrm{kin}}\) and a diffusion-limited velocity \(V_{\mathrm{diff}}\), i.e.,
\(V_{\mathrm{n}}^{- 1}\  \approx \ V_{\mathrm{kin}}^{- 1}\  + \ V_{\mathrm{diff}}^{- 1}\).
\(V_{\mathrm{kin}}\) can be estimated from Wilson--Frenkel-type expressions that
scale with the chemical-potential driving force, while \(V_{\mathrm{diff}}\)
scales with \(D\) times the ratio of the available undercooling to the
liquidus slope and the characteristic diffusion length ahead of the
interface. In concentrated CPA cocktails near the glass transition,
\(D\) becomes so small that \(V_{\mathrm{diff}}\) effectively vanishes and growth
is arrested even though the thermodynamic driving force remains
substantial. In moderately concentrated solutions at higher
temperatures, by contrast, \(D\) is large enough that interface kinetics
still play a significant role and growth can proceed rapidly once
nucleation has occurred.

Beyond one-dimensional planar growth, solute interactions with the
advancing front also determine the morphology of the ice phase. When the
solute boundary layer is sufficiently strong that the local liquidus
temperature ahead of the interface drops faster than the actual
temperature field, a region of so-called constitutional supercooling is
created. Under these conditions small perturbations of the interface are
amplified rather than damped, leading to cellular or dendritic ice
structures instead of a smooth front. In cryobiological samples this
translates into complex networks of extracellular ice that can
mechanically entrap cells and produce highly non-uniform solute
distributions, further complicating subsequent growth and warming
behavior.

After the initial growth stage, ice microstructure continues to evolve
even at fixed temperature through recrystallization. In broad terms,
recrystallization is a curvature-driven coarsening process in which
larger crystals grow at the expense of smaller ones. Because the
equilibrium melting temperature of a curved interface depends on its
radius via a Gibbs-Thomson relation, small, highly curved crystals have
a slightly higher chemical potential and lower melting temperature than
large, gently curved ones. This difference in chemical potential drives
net mass transfer from small to large crystals through
melting--recrystallization cycles, leading over time to a reduction in
total grain-boundary area and a progressive increase in the
characteristic crystal size.

Classical Ostwald ripening theories describe this coarsening in terms of
diffusion-limited transport of water through an intervening liquid or
glassy matrix, predicting a characteristic length scale that grows as a
power-law in time (often \(t^{1/3}\ \) for simple systems). In CPA-rich,
partially vitrified, or highly viscous solutions relevant to
cryopreservation, the effective diffusivity is strongly temperature
dependent, so that recrystallization is negligible at deep cryogenic
temperatures but can proceed rapidly when the sample is held or passes
slowly through warmer sub-zero ranges (for example, between roughly --20
\textdegree C and --5 \textdegree C). For cells and tissues, such coarsening can be
particularly damaging, because it transforms initially fine-grained,
relatively benign ice networks into larger crystals that can disrupt
membranes and extracellular matrices.

One of the major rationales for adding so-called ice recrystallization
inhibitors (IRIs) to cryopreservation media is precisely to slow or
arrest this curvature-driven coarsening. Natural anti-freeze proteins
and glycoproteins, as well as synthetic polymers such as certain
poly(vinyl alcohols) and related ampholytes, can adsorb selectively to
particular ice crystal faces, pinning step motion and reducing the
surface mobility needed for grain-boundary migration
\cite{r26,r27,r28,r29}. In continuum terms, these additives effectively
reduce the kinetic coefficient associated with interface motion without
necessarily changing the underlying thermodynamic driving force. The
result is a dramatic suppression of recrystallization over
experimentally relevant time scales, which can preserve a fine-grained
ice morphology during storage and through slow warming, thereby
mitigating mechanical and osmotic injury even when some ice is present.

\subsubsection{Heterogeneous nucleation}

In practical cryopreservation, ice almost never nucleates homogeneously
in the bulk solution. Instead, nucleation is overwhelmingly
\emph{heterogeneous}: it occurs on foreign surfaces such as container
walls, extracellular matrices, cellular membranes, or deliberately added
ice nucleating agents (INAs). From the perspective of classical
nucleation theory, these surfaces lower the free-energy barrier for
forming a critical nucleus, thereby shortening the induction time and
raising the probability of nucleation at comparatively mild supercooling
\cite{r30}.

The simplest and most widely used description of heterogeneous
nucleation is the spherical-cap model. Here, the ice embryo is assumed
to form as a cap on an ideal, flat substrate rather than as a full
sphere in the bulk liquid. The geometry is characterized by a contact
angle $\theta$ between the ice, the liquid, and the substrate. Under this
construction, the free-energy barrier for heterogeneous nucleation can
be written as a simple multiplicative correction to the homogeneous
barrier \cite{r30},

\begin{equation}
\Delta G_{\mathrm{het}}^{*} = f(\theta)\Delta G_{\hom}^{*}
\end{equation}

where \(\Delta G_{\hom}^{*}\) is the barrier derived in the homogeneous
case and \(f(\theta)\) is the geometric factor that accounts for the
reduced nucleus volume and surface area. For
\(\theta \rightarrow 180{^\circ}\) (perfectly non-wetting substrate),
(\(f(\theta) \rightarrow 1\)) and the surface has no effect. For
\(\theta \rightarrow 0{^\circ}\) (perfectly wetting substrate),
(\(f(\theta) \rightarrow 0\)) and the barrier may be reduced by many
orders of magnitude. Because the nucleation rate depends exponentially
on \(\Delta G^{*}/k_{\mathrm{B}}T\), even modest changes in \(\theta\) (and thus
in \(f\)) can transform nucleation from effectively impossible to nearly
instantaneous under otherwise identical conditions.

To connect this barrier reduction to observable nucleation rates, we may
simply replace \(\Delta G_{\hom}^{*}\) with
\(\Delta G_{\mathrm{het}}^{*}\) in the CNT rate expression, yielding a
heterogeneous nucleation rate \cite{r30}

\begin{equation}
J_{\mathrm{het}} = J_{0}\exp\left( - \frac{f(\theta)\Delta G_{\hom}^{*}}{k_{\mathrm{B}}T} \right)
\end{equation}

where \(J_{0}\) is again the kinetic prefactor. In the immersion mode
most relevant to cryobiology, nucleation is assumed to occur on active
sites that are fully wetted by the solution (e.g. a nucleator particle
or a patch of the container surface immersed in the liquid). The total
nucleation rate is then proportional to both the site-specific rate
(\(J_{\mathrm{het}}\)) and the density (or number) of such active sites, so that
both site strength (through $\theta$ or equivalent parameters) and site
abundance determine the macroscopic stability of a supercooled sample.

Real cryobiological systems, however, rarely present a single, idealized
contact angle, and thus are seldom adequately described by the spherical
cap interpretation of heterogeneous nucleation. Instead, a given surface
in contact with the liquid may host a spectrum of active sites, each
characterized by its own local barrier height and kinetics. Mineral
particles, polymer-coated chamber walls, proteins embedded in
extracellular matrices, and bacterial ice nucleating proteins all
provide distinct microenvironments for vicinal water molecules, and thus
distinct effective nucleation parameters \cite{r31}. Contemporary
descriptions of heterogeneous nucleation therefore supplement the
geometric contact-angle model with active-site distributions, in which
each site type is assigned a characteristic barrier (or contact angle)
and an areal density. The observed nucleation statistics then arise from
an extreme-value competition among these sites: nucleation almost always
occurs at the \emph{most active} site present, not at a hypothetical
average site (\textbf{Fig. 9c} shows family of extreme value
distributions describing active site variability). This framework
underlies the extreme-value statistics models increasingly used to
interpret cryobiological nucleation experiments and to quantify the
stability of supercooled systems \cite{r32}.

An alternative but complementary view, emphasized in more recent work by
Barahona \cite{r31,r33,r34}, focuses less on the static contact
angle and more on ice templating and the dynamics of vicinal water. In
this view, an active site is one that organizes nearby water into motifs
structurally compatible with ice (thereby lowering the entropic cost of
forming an ordered cluster) and/or modifies the local energy landscape
so as to reduce the work required to assemble a critical nucleus.
Crystalline substrates whose lattice spacings closely match those of
basal or prism planes of ice can act as nearly perfect templates,
strongly reducing the interfacial free energy and effectively lowering
\(\Delta G^{*}\). At the same time, adsorption of solutes or
macromolecules onto the substrate can slow local water mobility,
altering the kinetic prefactor, \(J_{0}\). Some authors therefore
introduce an explicit \emph{templating factor} that multiplies the
homogeneous rate by a function of both geometric matching and local
dynamical slowdown, rather than relying solely on a geometric
\(f(\theta)\). With the templating factor \(\zeta\) varying between 0
(no templating) and 1 (full templating), a nucleation site-specific,
surface area-normalized heterogeneous nucleation rate can be summarized
as

\begin{equation}
J_{\mathrm{het}} = \frac{Zf^{*}(\zeta)}{a_{0}}\exp\left( - \frac{\Delta G_{\mathrm{het}}^{*}(\zeta)}{k_{\mathrm{B}}T} \right)
\end{equation}

in which \(f^{*}(\zeta)\) is the attachment frequency factor, \(Z\) is
the Zeldovich factor, \(a_{0}\) is the molecular cross-sectional area of
water, and \(\Delta G_{\mathrm{het}}^{*}(\zeta)\) is the surface-dependent
nucleation barrier. This rate is plotted in \textbf{Fig. 8} below for
pure water.

\begin{figure}[h]
\centering
\includegraphics[width=0.67\textwidth]{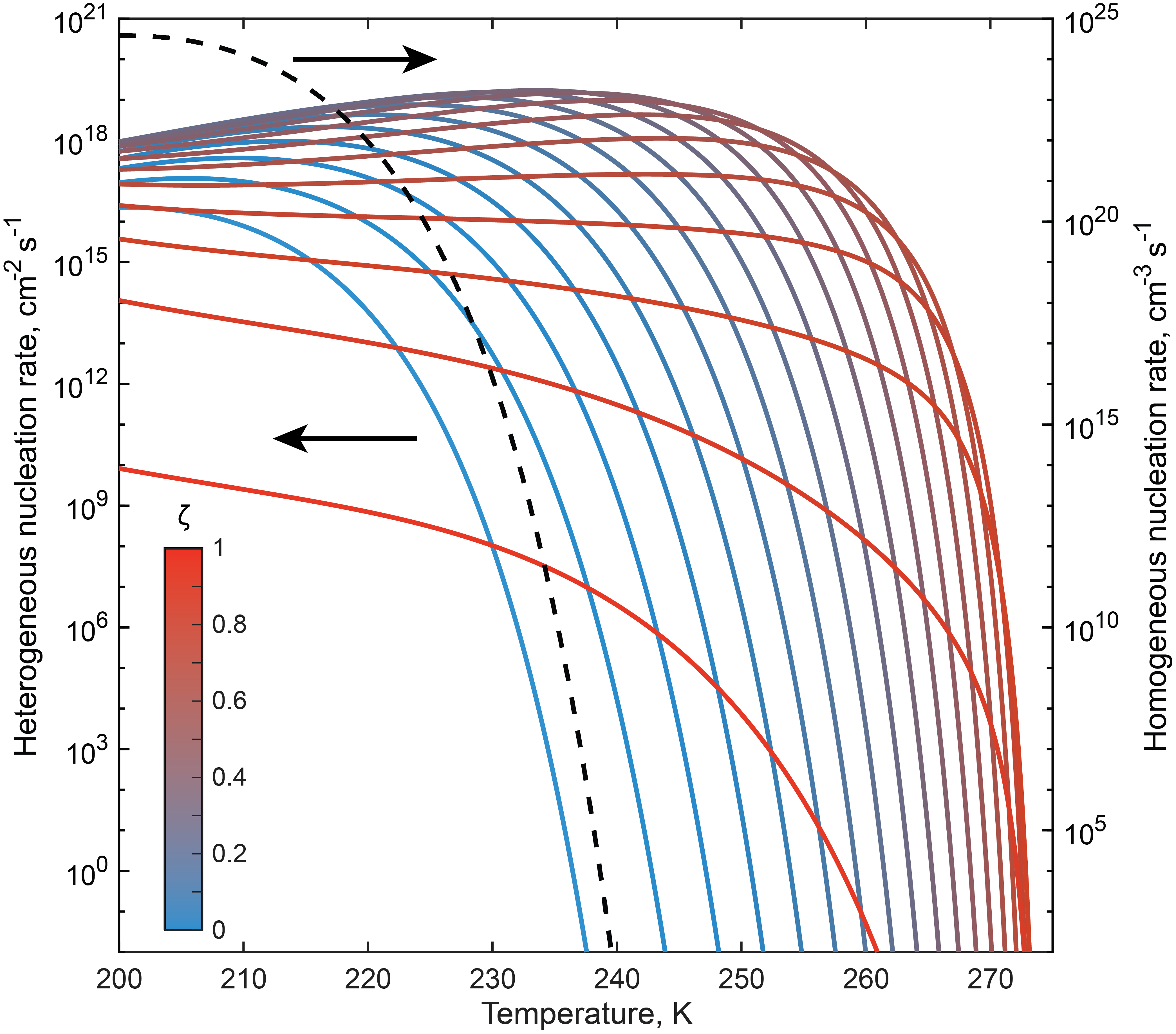}
\caption{Surface-scaling heterogeneous and volume-scaling
homogeneous ice nucleation rates as a function of temperature. Shown for pure water and various values of Barahona's ice ``templating factor'' \cite{r31}, which encodes the degree to which a foreign surface modifies the local free energy landscape so as to reduce the work required to assemble a critical nucleus.}
\label{fig:fig8}
\end{figure}

From a cryobiological design standpoint, heterogeneous nucleation is
both a nuisance and a tool. On one hand, unwanted active sites -
including microscopic scratches or contaminants on container walls,
undissolved particulates, or patches of gas-liquid interface -
dramatically raise nucleation temperatures and undermine attempts to
maintain supercooling or achieve vitrification. Mitigation strategies
therefore focus on minimizing, passivating, or isolating such sites:
polishing or coating surfaces, eliminating gas bubbles, overlaying
solutions with immiscible liquids, or confining samples in smooth,
rigid, air-free isochoric chambers (\emph{see} \textbf{Section
3.2}). On the other hand, for ice seeding and partial freezing
protocols, the goal is, precisely to \emph{create} highly active
heterogeneous sites in carefully chosen locations (e.g. in the
extracellular space or in the perfusate), so that nucleation occurs
reproducibly at modest supercooling and in mechanically benign regions.
Engineered ice nucleating agents, from mineral particles to synthetic
macromolecules and bacterial ice nucleation proteins, can thus be seen
as deliberate manipulations of the heterogeneous nucleation landscape,
tuning both the barrier (via geometry and templating) and the site
density to produce desired nucleation statistics.

In summary, heterogeneous nucleation extends classical nucleation theory
by incorporating the effects of real surfaces and interfaces into the
free-energy barrier and kinetic prefactor. As such, whether a given
cryopreservation protocol succeeds in stabilizing a supercooled or
vitrified state, or in achieving a well-timed ice seeding event, may
depend less on the homogeneous properties of the bulk solution and more
on the nature, distribution, and management of these heterogeneous
nucleation sites.

\subsubsection{Statistical aspects of ice nucleation}

While classical nucleation theory describes nucleation as a
\emph{continuous} process in time via the nucleation rate, \emph{J,} the
formation of a single nucleus remains a discrete process: long,
seemingly quiescent periods with no events are punctuated by individual
instances of critical nucleus formation. Despite the conceptual
prevalence of CNT, what the cryobiologist actually \emph{observes} in
most experiments and protocols are these discrete nucleation outcomes,
e.g. ``frozen'' vs. ``unfrozen'', recorded at particular times or
temperatures. Connecting these discrete and continuous views of
nucleation enables marriage of laboratory outcomes to broader kinetic
understanding of the kinetic phenomena at play, and requires treating
nucleation as a purely probabilistic process.

This may be achieved by use of Poisson statistics, which describe the
probabilistic nature of memoryless stochastic events \cite{r35}.
Within this framework, for a system at temperature \(T\) with an
effective nucleation rate \(J_{\mathrm{sys}}(T)\), the expected number of
nucleation events accumulated over a time interval
\(\lbrack 0,t\rbrack\) is \(J_{\mathrm{sys}}(T) \cdot t\). Assuming that each
possible nucleus forms independently, the probability of observing no
nucleation events in that interval is

\begin{equation}
S(T,t) = \exp\left( - J_{\mathrm{sys}}(T)t \right)
\end{equation}

which is the survival probability of the metastable liquid. The
probability of at least \emph{one} nucleation event having occurred by
time, \(t\), is then \(P(T,t)\  = \ 1\  - \ S(T,t)\). Formally, this is
the cumulative distribution function of a homogeneous Poisson process
with rate \(J_{\mathrm{sys}}\) (see example distributions in \textbf{Fig. 9a}).
The corresponding probability density function for the waiting time to
the first event is

\begin{equation}
p(T,t) = \frac{\partial P}{\partial t}\  = \ J_{\mathrm{sys}}(T)\exp\left( - J_{\mathrm{sys}}(T)t \right)
\end{equation}

which is an exponential distribution. The characteristic time until
nucleation (the mean waiting time, often referred to as the induction
time) is therefore the average time \(\left\langle t \right\rangle\):

\begin{equation}
t_{\mathrm{nuc}} = \frac{1}{J_{\mathrm{sys}}(T)}
\end{equation}

More generally, if one wishes to guarantee that the probability of
nucleation does not exceed some small failure probability \(p^{*}\)
during a hold at temperature, \(T\), the maximum allowable hold time is

\begin{equation}
t_{p^{*}}(T) = - \frac{\ln\left( 1 - p^{*} \right)}{J_{\mathrm{sys}}(T)}
\end{equation}

For example, (\(p^{*} = 0.01\)) corresponds to a ``1\% risk'' induction
time; \(p^{*} = 10^{- 4}\) corresponds to the ``one-in-ten-thousand''
threshold often used in high-reliability engineering contexts. In most
cryobiological experiments and protocols, however, the system is not
held at a single temperature. Instead, the sample is cooled or warmed
according to some protocol, so both \(T\) and \(J_{\mathrm{sys}}\) vary with
time. For a general time-dependent temperature history \(T(t)\), the
survival probability becomes

\begin{equation}
S\left( T(t) \right) = \exp\left( - \int_{0}^{\mathrm{t}}{J_{\mathrm{sys}}\left( T(\tau) \right)d\tau} \right)
\end{equation}

This is the defining form of a non-homogeneous Poisson process. A
particularly important special case for cryopreservation is
constant-rate cooling. If the system is cooled from the equilibrium
melting point, \(T_{\mathrm{m}}\), at a constant rate, \(R = |dT/dt|\), then time
and temperature are related by \(T(t) = T_{\mathrm{m}} - Rt,\ \)and the integral
in the survival function can be re-expressed as an integral over
temperature:

\begin{equation}
S_{\mathrm{CCR}}(T) = \exp\left( - \frac{1}{R}\int_{T_{\mathrm{m}}}^{\mathrm{T}}{J_{\mathrm{sys}}(\theta)d\theta} \right)
\end{equation}

This is the non-homogeneous Poisson distribution in temperature, used in
constant-cooling-rate experiments to extract nucleation rates from
distributions of observed freezing temperatures. Differentiating with
respect to temperature yields the probability density for nucleation to
occur at temperature, \(T\):

\begin{equation}
p_{\mathrm{CCR}}(T) = \frac{\partial P_{\mathrm{CCR}}}{\partial T} = \frac{J_{\mathrm{sys}}(T)}{R}\exp\left( - \frac{1}{R}\int_{T_{\mathrm{m}}}^{\mathrm{T}}{J_{\mathrm{sys}}(\theta)d\theta} \right)
\end{equation}

An example cumulative distribution function and probability density
function are shown in \textbf{Fig. 9b}.

The characteristic nucleation temperature, \(T_{\mathrm{nuc}}\), can then be
defined in several equivalent ways; for example as the mean of this
distribution,

\begin{equation}
T_{\mathrm{nuc}} = \int_{0}^{T_{\mathrm{m}}}{Tp_{\mathrm{CCR}}(T)dT}
\end{equation}

or, more practically, as its mode or median. In systems where
\(J_{\mathrm{sys}}\) varies extremely rapidly with undercooling (as is typical in
CNT), \(p_{\mathrm{CCR}}(T)\) becomes sharply peaked, so that the ``nucleation
temperature'' is very well-defined even though the underlying process is
stochastic.

Two additional features of the Poisson description are particularly
important for cryobiology. Firstly, in the simplest homogeneous case,
the system nucleation rate is proportional to the system volume, so
larger samples have proportionally higher nucleation rates and
correspondingly shorter induction times. This makes it much easier to
maintain deep supercooling or vitrification in droplets or thin films
than in organs or large tissues, even if the underlying solution
thermodynamics and interfacial properties are identical. More refined
models replace the single volume, \(V\), with a sum over compartments
(e.g. intracellular vs. extracellular, or separate regions of an organ),
each with its own local \(J\) and effective volume.

Secondly, in heterogeneous nucleation, the system often contains a
spectrum of active sites, each with its own local nucleation rate
\(J_{i}(T)\). If each site is modeled as an independent Poisson process,
the system-level survival probability is product of each independent
survival probability, and the system level nucleation rate is a sum of
the individual active site nucleation rates. When the \(J_{i}\) span
many orders of magnitude, nucleation almost always occurs at the single
most active site present, making the observed distribution of induction
times or nucleation temperatures an extreme-value statistic of the
underlying site population rather than a simple reflection of an average
rate. This is the basis for active-site and extreme-value models that
infer the strength and abundance of heterogeneous nucleation sites from
experimental nucleation statistics in supercooled cryobiological systems
\cite{r32}.

\begin{figure}[b]
\centering
\includegraphics[width=1.0\textwidth]{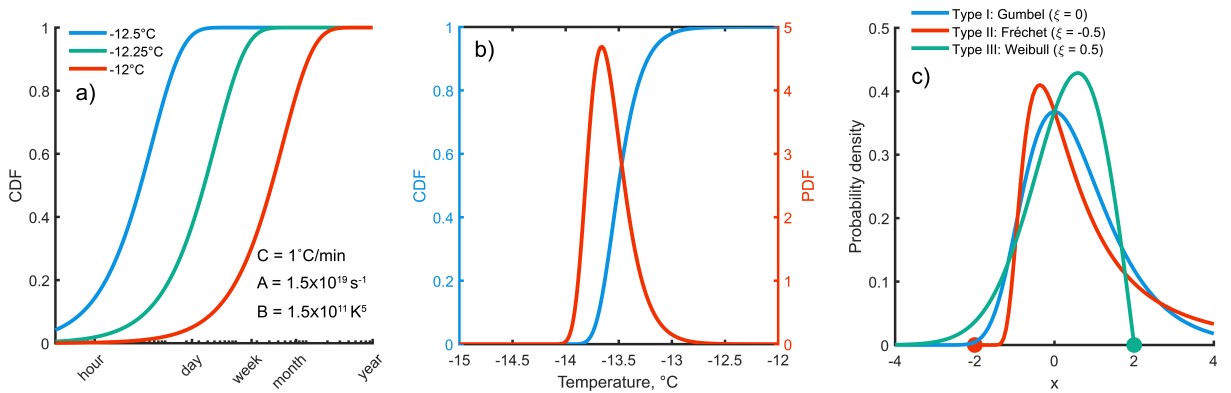}
\caption{Nucleation statistics a) Cumulative distribution
function for isothermal supercooling (i.e., freezing probability). b)
Survivor curve (blue/right) and probability density function for
constant cooling rate. c) Extreme value distributions which describe
distribution of heterogeneous nucleation active sites. Adapted with
permission from Consiglio et al. \cite{r32}.}
\label{fig:fig9}
\end{figure}

In practice, these statistical relations provide the bridge between
measured freezing temperatures and underlying nucleation physics, and
they supply the most directly useful design quantities for
cryopreservation. The statistical metrics summarized here underpin many
of the emerging cryopreservation technologies discussed below. They
enable rational selection of cooling and warming profiles, storage
temperatures, and sample geometries that keep the cumulative nucleation
probability acceptably low (for supercooling and vitrification) or,
conversely, guarantee prompt nucleation at desired locations and times
(for ice seeding and partial freezing). We will utilize these statistics
in Case Study 2 later in this work.

\subsection{Heat and mass transport}

Now having formulated a description of the thermodynamic equilibrium
states that cryobiological water may occupy and the molecular-level
kinetic process that facilitates transition between these states, we
arrive at the final physics that connect the practitioners and protocols
of cryobiology to these phenomena---heat and mass transfer.

In practice, the cryobiologist has two principal means by which to
control the outcome of a cryopreservation process, in terms of the final
states attained and the phase transition route followed to arrive at
those states: the first is chemical design of the cryobiological
solution employed; the second is choice of thermal and mass transport
history (cooling rates, warming rates, CPA loading rates, etc.).
Historically, cryobiology literature has addressed the latter much more
systematically than the former, and (for better or worse) much of the
philosophy of cryopreservation protocol design today focuses on
engineering of thermal and mass flow histories. Accordingly, there are
many excellent reference works addressing cryobiological heat and mass
transfer, and we will refer the reader to them rather than discussing
comprehensively herein. We particularly recommend the formidable recent
review by Zhao and colleagues \cite{r36}, a salient chapter of
the recent cryopreservation-focused \emph{Annual Review of Heat
Transfer} \cite{r37}, and the Chapters of this book authored by Higgins and colleagues and Shu and colleagues.

Instead, we will here highlight the oft-overlooked intersections of heat
and mass transport with the equilibrium thermodynamics and phase change
kinetics at play, with a particular focus on relevant transport
properties affecting those frameworks.

\subsubsection{Heat transport - key concepts}

The temperature history of a sample is one of the most important
manipulable aspects of a cryopreservation protocol, underlying Mazur's
famous two-factor hypothesis \cite{r21}; the process of
vitrification \cite{r38}; and many other cryobiology essentials.
However, the effect of this temperature history is nearly meaningless
without its connection to the thermodynamics and kinetics at play, which
it affects through the following broad aspects:

\subsubsection*{Temperature-time dependence of ice nucleation and growth}

As described in the preceding section, the nucleation rate and
probabilistic induction time of ice formation are functions of thermal
history, a fact which may be leveraged in several ways. A few examples
include: 1) During the slow freezing of cells, where extracellular ice
nucleation is desired, the probable induction time of nucleation may be
calculated as a function of cooling profile for the extracellular and
intracellular compartments separately (even distinguishing only by
compartmental volume), and a cooling profile that ensures a
significantly smaller induction time for the extracellular compartment
may be chosen. This is the principle behind ``interrupted freezing''
protocols \cite{r39}. 2) During the vitrification of any
biologic, the cooling profile required to reach the glass transition
temperature in less than one induction time (corresponding to perfect
vitrification, and the most rigorous form of the ``critical cooling
rate'') may be estimated. 3) During supercooling, a cooling profile and
storage temperature may be calculated such that the induction time for
ice nucleation is much longer than the desired storage period. The
cooling profiles required will be functions of both the underlying
solution thermodynamics and the heterogeneous surfaces with which the
system is in contact \cite{r32,r40}.

\subsubsection*{Temperature-time dependence of metabolism \& toxicity}

The core mechanism enabling cryopreservation is the arrest of
metabolism, which scales exponentially with temperature (\emph{see}
\textbf{Fig. 15} in Case Study 2 below). The toxicity of
cryoprotectants, the rate of which is driven both by the rate of
metabolism and by the rate of various potential biochemical reactions
and phase change processes, is likewise exponentially slowed upon
cooling from normothermia. As a result, the temperature history of a
given protocol is inextricably linked both to the plausible period of
preservation achievable (based strictly on metabolism \cite{r41})
and to the toxicity accumulated by the sample. Thus, if the functional
form of the temperature dependence of metabolism is known, metabolism
arrest (i.e. preservation benefit) may thus be mathematically coupled to
cryopreservation physics, as demonstrated by Consiglio et al. in the
case of organ-scale supercooling \cite{r42}. Likewise, if the
functional form of the temperature-dependence of toxicity rate is known,
accumulated toxicity may be mathematically coupled to cryopreservation
physics, as demonstrated by Benson and colleagues in the case of
cellular vitrification \cite{r43,r44,r45}.

\subsubsection*{Temperature-time dependence of thermomechanical stresses}

Though a topic that has received comparatively minimal treatment
historically, growing interest in cryopreservation of large biological
samples (e.g. organs and organisms) has driven recent interest in the
calculation and avoidance of thermal stresses that may accumulate due to
either phase change or differential thermal contraction across a large,
low-thermal-diffusivity sample during cooling. The first physics-based
investigations of cryopreservation-related thermal stresses pertained to
those developed during freezing \cite{r46}, wherein the dynamic
solidification front may drive significant local stresses. More
recently, focus has turned to the considerable challenge of glass
cracking during vitrification, which may manifest at even the microliter
sample scale \cite{r47}, and which proves especially challenging
in the vitrification of macroscale tissues and organs \cite{r48,r49}. Cryopreservation thermomechanics presents a particularly rich
intersection of solution thermodynamics and transport phenomena, with
recent work demonstrating sharp dependences both on thermodynamic
properties of the system (glass transition temperature, thermal
expansion coefficient \cite{r50}), transport properties (thermal
conductivity, viscosity), and precise sample geometry and localized
temperature history \cite{r51}.

\subsubsection*{Temperature-dependent thermophysical properties}

We note finally that heat transport and the thermal history of a sample
are coupled to every piece of physics described heretofore by the
temperature-dependence of both thermodynamic properties (i.e. first or
higher-order derivatives of the free energy, such as density, entropy,
chemical potential, thermal expansion coefficient, heat capacity, etc.)
and transport properties (thermal conductivity, viscosity, etc.).
However, while this coupling is well appreciated, there remains an acute
lack of low-temperature thermophysical property data for most solutions
of interest to cryobiology (and a similar lack of predictive tools to
cover this data gap), presenting a meaningful challenge to empirical
accurate thermophysical description of various cryopreservation
processes.

Gratefully, recent works indicate a growing, field-level effort to
address this critical need for low-temperature and cryobiology-relevant
thermophysical data. We point the reader to several key datasets
commonly used in recent cryobiological modeling: that assembled by Rabin
and colleagues, relevant to thermomechanics analysis of DMSO-based
vitrification media \cite{r52,r53}; by Bischof and colleagues,
relevant to critical cooling and warming rates for vitrification
\cite{r54}; by Elliott and colleagues, relevant to melting point
depression of various cryobiological solutions \cite{r10}; by
Powell-Palm and colleagues, relevant to the viscosity of many-component
aqueous solutions \cite{r55}; and by Consiglio and colleagues,
relevant to high-pressure phase equilibria \cite{r19,r56,r57}.

\subsubsection{Mass transport - key concepts}

Mass transport affects the practice and outcomes of cryobiology through
various important processes and effects, both molecular (e.g. \emph{via}
the diffusivity of water in solution, which critically affects the
nucleation process \cite{r58}) and biological (e.g. \emph{via}
the delivery of cryoprotectants across cell membranes
\cite{r58}). Gratefully, a thorough treatment of cryobiological mass transport is provided in the Chapter of this book authored by Higgins and colleagues, covering delivery of cryoprotectants into cells, tissues, and organs (by both diffusion and perfusion), the fundamental role of osmotic stress in designing cryopreservation protocols, etc. 
For additional reading on the specific interplay between cryobiological
mass transport and ice nucleation and growth processes, see these
various references \cite{r58,r60,r61,r62,r63,r64,r65,r66,r67}.

\section{Emerging techniques for cryopreservation of complex
biomaterials}
Armed now with sufficiently diverse thermodynamic theory to analyze the
physical aspects of nigh-any cryopreservation protocol, we proceed to
review emergent cryopreservation approaches through the lens of this
theory. Our strategy in the text to follow is to provide an accessibly
comprehensive survey of new or underexplored techniques and philosophies
(and especially those for which alternative review works do not yet
exist); to aid the reader in identifying and understanding the relevant
physical theory guiding these approaches; and, where possible, to
demonstrate the power of the theoretical tools articulated in
\textbf{Section 2} to enhance both interpretation and practice
of these protocols.

For consistency with our theoretical approach, we organize the emergent
technologies discussed herein by the theoretical aspect---equilibrium
thermodynamics, kinetics, or transport---that most principally drives
their design or practice. We also include a final section dedicated to
vitrification, as a prototypical technique drawing equally from all
three aspects. For each section, we will review broadly the relevant
techniques, then choose one on which to perform an illustrative case
study in the application of relevant theory to meaningful,
technique-specific analysis.

\subsection{Equilibrium techniques}

While all cryopreservation protocols are beholden to relevant phase
equilibria (e.g. melting points, eutectic points) and thermodynamic
properties (e.g. heat capacity, thermal expansion coefficient), only a
particular subset are largely decoupled from the short time-scale
phenomena associated with nucleation and transport processes. For these
privileged few techniques, equilibrium solution thermodynamics dominate
protocol design and practice, and they thus lend themselves to
relatively simple (and powerful) first-order analyses.

\subsubsection{Liquidus tracking}

The technique perhaps most emblematic of the equilibrium-driven approach
is \emph{liquidus tracking}. As its name suggests, this approach aims to
simultaneously introduce a cryoprotectant solution into a sample whilst
reducing its temperature, such that the sample follows (in both
composition and temperature) the liquidus curve of the solution
(\textbf{Fig. 10}). By maintaining the sample at the lowest temperature
at which the possibility of freezing may be systematically avoided (the
liquidus temperature), this process minimizes cryoprotectant toxicity
during loading to the highest degree achievable in a globally stable
state, and provides the fastest means of cooling an organ to sub-0 \textdegree C
temperatures (starting from a CPA-unloaded, supra-0 \textdegree C state) without
accepting \emph{any} degree of ice formation risk.

First proposed by Farrant in 1965 \cite{r68} and powerfully
realized by Elford \& Walter in 1972 \cite{r69}, liquidus
tracking in the modern era has been nigh-exclusively implemented as a
CPA-loading strategy, as opposed to a final storage method. In practice,
a sample is shuttled along an equilibrium concentration-temperature path
towards a target final CPA composition, from which a final cooling step
(typically to beneath the glass transition temperature) may then be
performed with minimal risk of freezing (even while departing then from
global equilibrium). This approach was formalized by Pegg and colleagues
in the preservation of cartilage \cite{r70}, wherein controlled
stepwise introduction of DMSO and cooling of the sample, up to a
terminal concentration of \textasciitilde60\% by mass and a terminal
loading temperature of --70 \textdegree C (approximately 10 \textdegree C above the according
liquidus temperature, as a margin of safety), permitted complete
avoidance of ice formation during CPA equilibration, and achievement of
tissue vitrification thereafter.

Pegg's approach, explored further in cartilage by his group
\cite{r71}, and in encapsulated 3D cell constructs by Fuller and
colleagues \cite{r72}, relied largely on preliminary empirical
quantification of both liquidus temperatures as a function of
composition, the rate or degree of CPA penetration at each step, and
other protocol parameters. More recent work has advanced the technique
towards its logical optimum, leveraging the imminent modelability of
liquidus curves to predictively design optimal temperature-concentration
protocols. In perhaps the most rigorous example of rationally-designed
liquidus tracking to date, Shardt, Elliott and colleagues used solution
theory (in particular multi-component osmotic-virial equations, as
described in \textbf{Section 2.1}) to power calculate both
liquidus coordinates and permeation driving forces for a multi-component
CPA cocktail in articular cartilage, systematically reducing the
time-toxicity profile of cryoprotectant loading to precede vitrification
\cite{r59}.

Finally, we note that while liquidus tracking has to date been applied
only to diffusively loaded cell and tissue constructs, Wowk and Taylor
et al. have proposed its use in perfusion-loaded vascular systems,
including whole organs \cite{r73,r74}, and as a final storage
technique independent of vitrification. In light of the following, we
find this to be a promising route for future exploration: 1) recent
studies in sub-0 \textdegree C whole-organ preservation have successfully utilized
the \emph{supercooled} liquid state for storage \cite{r3,r75,r76,r77};
2) multi-pump, multi-thermic perfusion loading systems are increasingly
becoming standard tools in organ cryopreservation; 3) metastable storage
techniques using penetrating cryoprotectants only achieve
\textasciitilde1--2 \textdegree C supercooling, introducing freeze-risk that may
outweigh the benefit of the marginally colder temperatures enabled; and
4) a wide diversity of new cryoprotectants show minimal toxicity at
concentrations (3--6 molal \cite{r78,r79}) that may enable
significantly colder preservation (\textasciitilde{} --10 to --20 \textdegree C) than
has yet been achieved in ice free preservation of whole organs.

\begin{figure}[t]
\centering
\includegraphics[width=0.67\textwidth]{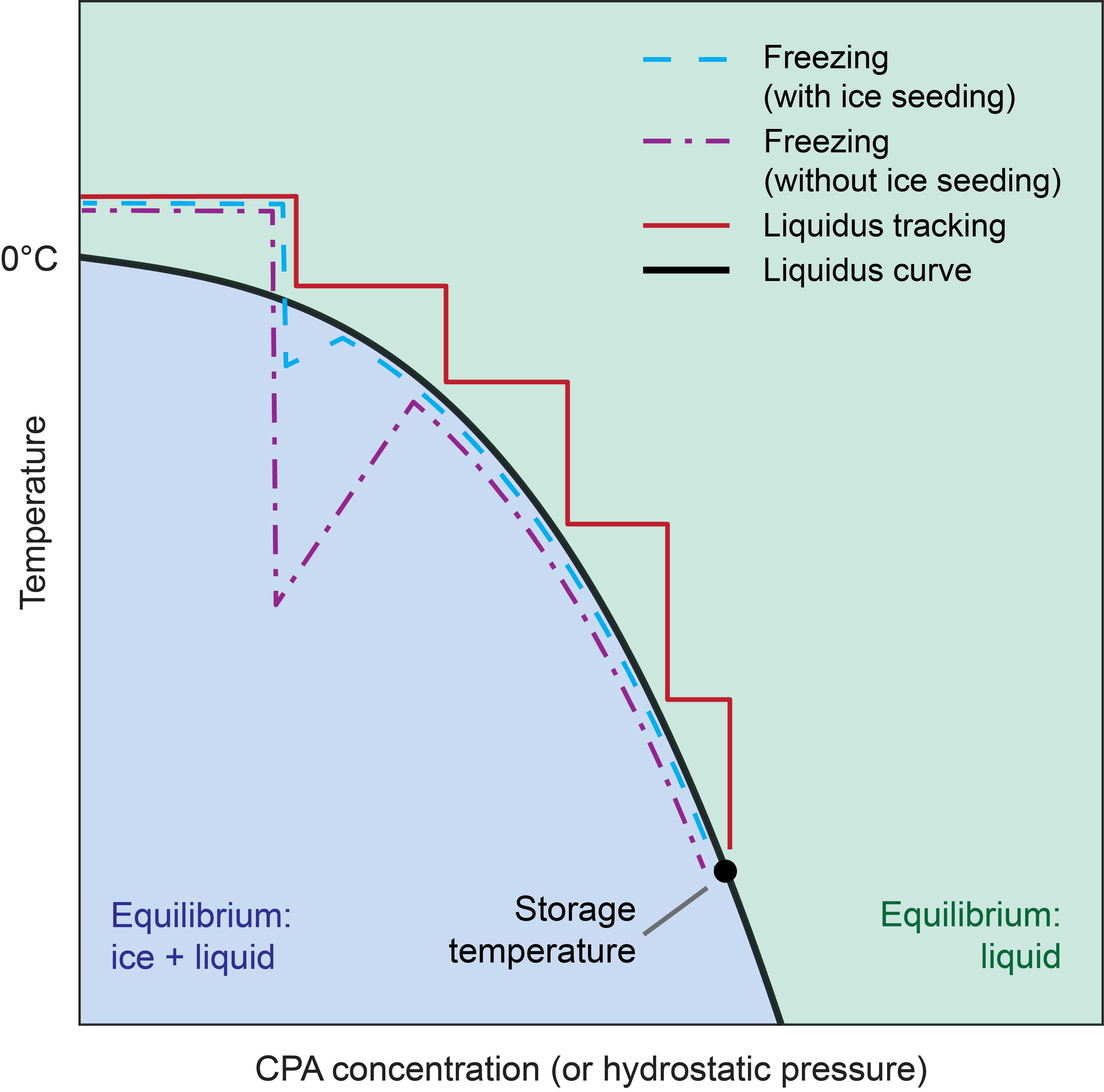}
\caption{Conceptual illustration of the thermodynamic paths
followed by various equilibrium cryopreservation techniques. The
liquidus curve, defining the melting point of ice as a function of CPA
concentration or pressure, is given by the thick black line. In liquidus
tracking (thin red line), a biological sample is simultaneously
CPA-loaded and cooled, such that it follows the liquidus curve. This
approach enables CPA-loading and sample storage at the lowest possible
temperatures (and accordingly toxicities) at which the ice formation is
not a risk. In partial freezing (dashed lines), a sample is loaded with
an initial concentration of CPA, cooled below the liquidus, and allowed
to form a two-phase liquid-ice solution. The amount of ice that will
form at a given temperature is prescribed by the lever rule, and
freeze-driven solute rejection will ripen the remaining liquid to the
liquidus concentration at that same temperature. Seeding ice can
minimize the osmotic shock accompanying this solute rejection (and
mechanical damage from crystallization) upon initial nucleation from the
supercooled liquid. Isochoric freezing follows an analogous
thermodynamic path to partial freezing, but in either
temperature-pressure space (if the liquid medium in the chamber is pure
water, as is commonly used), or in 3D temperature-pressure-concentration
space. We note that in partial freezing however ice forms \emph{within}
a biological sample, while in isochoric freezing ice forms
\emph{outside} of the sample.}
\label{fig:fig10}
\end{figure}

\subsubsection{Partial freezing}

While liquidus tracking uses thermodynamic equilibrium to ensure
complete avoidance of ice in a biological sample, the same liquidus
equilibria may be leveraged to \emph{allow} freezing in the sample but
systemically limit its extent. This approach, demonstrated by Ishine,
Rubinsky \& Lee in the 1990s \cite{r80,r81} and later named
``partial freezing'' by Tessier, Toner, Uygun and colleagues
\cite{r82,r83,r84}, occupies the equilibrium corridor between liquidus
tracking and classical slow-freezing, targeting temperature-composition
coordinates that situate the sample beneath its liquidus temperature but
above the eutectic temperature, rendering a thermodynamically-stable
two-phase liquid-ice equilibrium within, with the amount of ice
prescribed by the lever rule (\textbf{Fig. 10}).

Use of partial freezing, as opposed to liquidus tracking, is motivated
by the deeper temperatures of preservation achievable per-unit
cryoprotectant loaded, at the cost of bulk ice formation. However, we
note that \emph{at equilibrium}, due to solute rejection, the
cryoprotectant concentrations \emph{in the portion of the system which
remains liquid} will be precisely identical to those that would enable
liquidus tracking at the same temperature, raising intriguing questions
as to whether the benefits of reduced load concentrations outweigh the
dangers of ice formation.

This technique has been developed principally for the preservation of
vascular organs, in which a cryoprotectant solution may be delivered via
machine perfusion or manual flush, and this ice formation may be
strategically initiated in the vascular space, where it is hypothesized
that it may nucleate, grow, and reject solutes with minimal mechanical
interference. This initiation has been observed to accompany risk of
damage to the endothelium \cite{r80,r82}, both by mechanical
mechanisms related to ice growth and biochemical mechanisms related to
solute ripening. However, preliminary \emph{ex vivo} evaluations of
mammalian livers \cite{r82,r84} and kidneys \cite{r85}
nonetheless appear promising.

In order to more deeply interrogate the thermodynamic and physical
implications of partial freezing, and to demonstrate a prototypical
thermodynamic analysis of a two-phase liquid-ice system, we conduct a
Case Study of the partial freezing technique described by Tessier et al.
at the end of this section.

\subsubsection{Isochoric freezing}

The vast majority of cryopreservation protocols are performed under a
constant (typically atmospheric) pressure, wherein the volume of the
sample may freely expand or contract in response to phase change,
temperature change, composition change, or any combination thereof. In
2005, Rubinsky and colleagues proposed the opposite: to constrain the
\emph{volume} of a sample, thereby allowing its \emph{pressure} to vary
in response to these same changes \cite{r18}. This process,
called isochoric (constant volume) or confined freezing, seeks to
leverage hydrostatic pressure passively generated by the cooling process
to depress the melting point of water, thereby protecting a biological
sample from ice formation at temperatures beneath its
atmospheric-pressure melting point.

\begin{figure}[!b]
\centering
\includegraphics[width=0.6\textwidth]{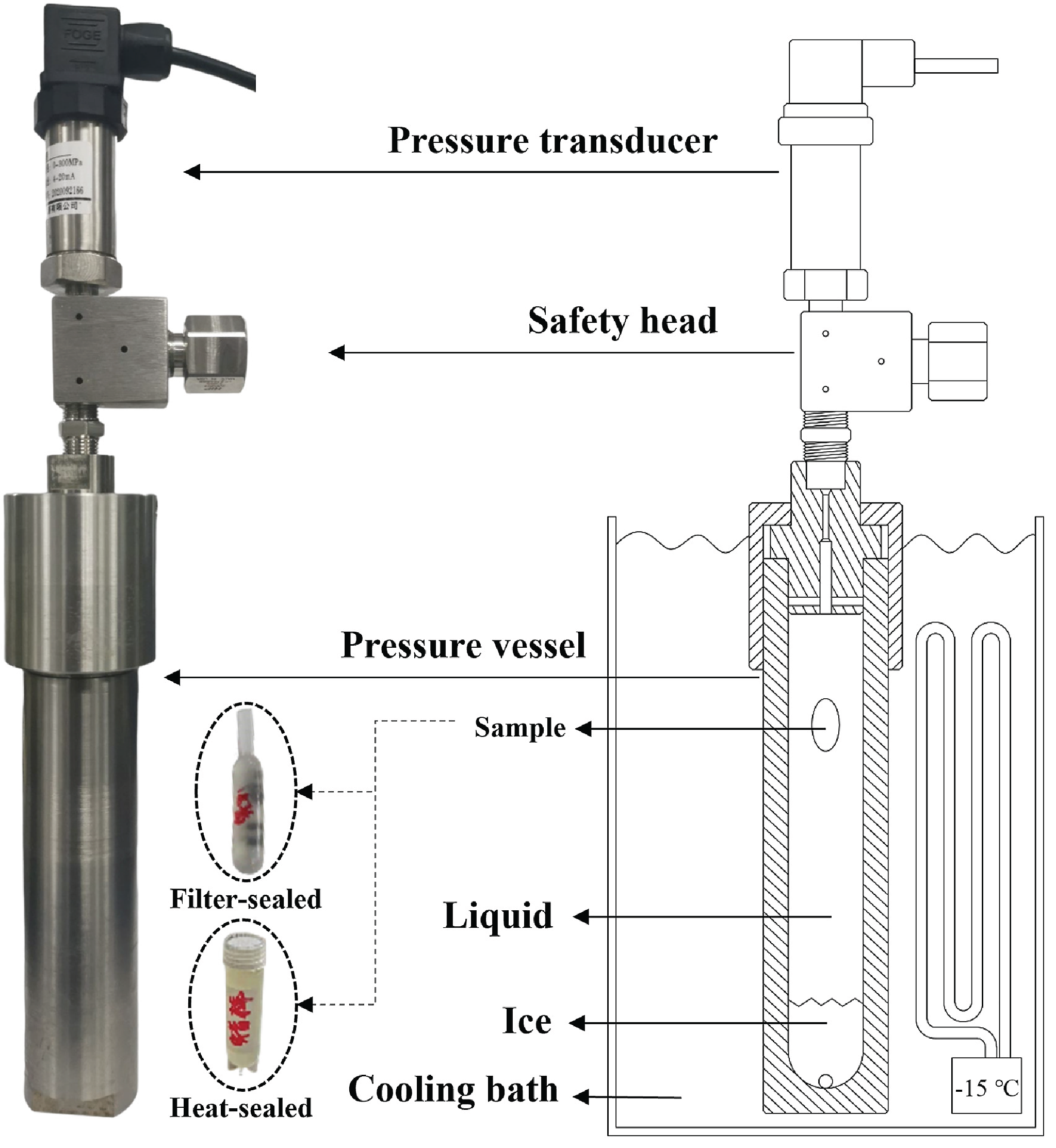}
\caption{Isochoric freezing chamber and schematic (adapted with
permission from Lyu et al. \cite{r86}).}
\label{fig:fig11}
\end{figure}

In practice, isochoric freezing typically involves submerging a
biological sample in an aqueous medium within a rigid, high-strength
chamber; sealing the chamber absent any air within; cooling the chamber
to a sub-0 \textdegree C temperature; and initiating ice nucleation at a position
in the chamber distant from the biologic (see \textbf{Fig. 11}). The
growth and expansion of this ice will then, given its rigid confinement
and isolation from the atmosphere, drive pressurization (on the order of
10--210 MPa) of the remaining liquid in the system (which contains the
biologic), the melting point of which is in turn depressed until the ice
and liquid achieve a two-phase equilibrium, as prescribed by the
pressure-temperature-composition (p-T-x) liquidus curve.

As such, isochoric freezing represents the same thermodynamic principle
as partial freezing: ice is deliberately nucleated and allowed to
``ripen'' the remaining liquid (whether in concentration, in pressure,
or in both) until thermodynamic equilibrium is reached. However, two key
differences distinguish these techniques. First, in isochoric freezing,
the limited formation of ice within the system occurs \emph{outside} the
biological sample, as opposed to within it, thereby providing ice-free
preservation at sub-0 \textdegree C temperatures. Second, isochoric freezing may be
practiced with or without cryoprotectants, as pressure depresses the
melting point of liquid water universally. As such, cryoprotectants in
isochoric freezing carry only a portion of the burden of ice protection,
enabling ice-free preservation at lower cryoprotectant concentrations at
a given temperature \cite{r87,r88}. However, just as the
temperature advantages of partial freezing are weighed against the
potential dangers of ice formation, so the advantages of isochoric
freezing are weighed against the potential dangers of pressure.

For a deeper discussion of the principles of isochoric freezing, in
addition to a comprehensive summary of its application to multi-scale
biologics ranging from mammalian hearts to fresh food products, we refer
the reader to our recent review of this topic \cite{r19}.

\subsubsection{Case study 1: Thermodynamic analysis of chemical evolution
during partial freezing}

Partial freezing, in the limit of prompt nucleation and slow cooling
rates, lends itself to straightforward analysis with the toolkit of
equilibrium solution thermodynamics. In this Case Study, we will
demonstrate simple applications of this toolkit to the interrogation of
a partial freezing solution and protocol reported by Tessier et al.
\cite{r1}, and in doing so illustrate routes towards rational
design of partial freezing solutions and protocols. A general process is
as follows:

1. Select a multi-solute solution thermodynamics model to describe the
chemical potential of the partial freezing solution of interest and
utilize it to calculate the melting point.

2. Acknowledging that ice forms a pure solid phase, effectively removing
water the liquid solution, calculate the partial freezing-relevant
liquidus curve by iteratively decrementing the concentration of water
whilst holding all other component concentrations constant, and
re-calculating the melting point for each decrement.

3. Use this liquidus curve and the lever rule to calculate the amount of
ice that will form at a given sub-freezing temperature, as a fraction of
the total system mass or volume. Calculate the according concentrations
of total or individual solutes by removing the calculated amount of ice
from the initial liquid water content of the solution and re-calculating
the concentrations.

4. Use these insights to choose an appropriate storage temperature,
based on desired limits of ice formation, terminal concentrations of
cryoprotectants or electrolytes in the unfrozen liquid phase, etc.
Alternatively, use the steps above to \emph{design} a solution that
provides optimal values of the aforementioned.

\begin{figure}[h]
\centering
\includegraphics[width=0.9\textwidth]{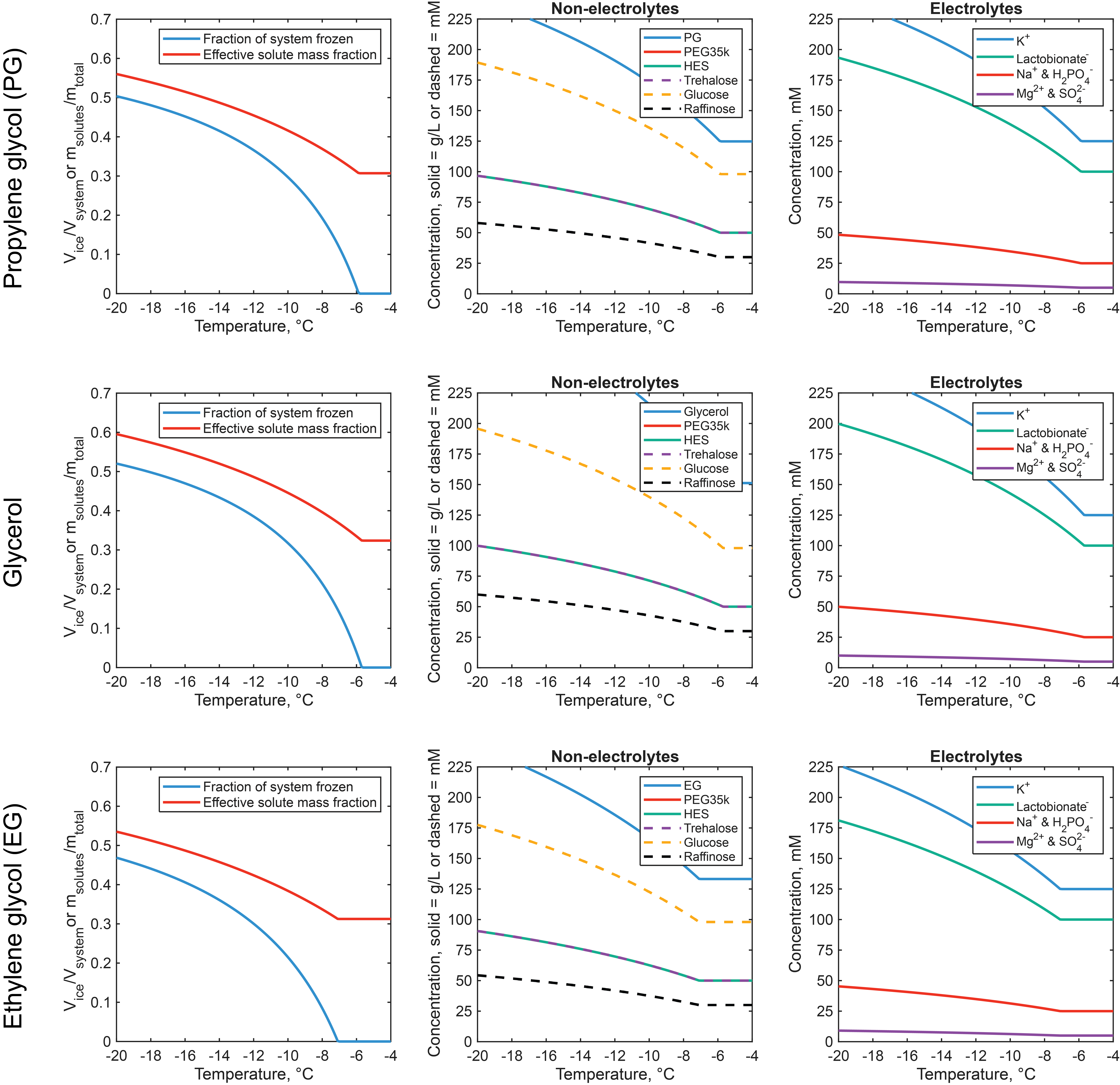}
\caption{Thermodynamic analysis of a partial freezing solution.
We analyze the compositions reported by Tessier et al. \cite{r1},
which vary only by the choice of penetrating cryoprotectant. Results for each cryoprotectant are organized by row as labeled. The first column gives the ice fraction and the effective solute mass fraction in the remaining liquid phase at thermodynamic equilibrium, as a function of temperature. The second and third columns give the freeze-driven evolution of the concentrations of each chemical component as a function of temperature.}
\label{fig:fig12}
\end{figure}

Using the partial freezing solution composition provided by Tessier et
al., we show prototypical outputs of this process in \textbf{Fig. 12}.
For each of the three variants investigated by the authors (which differ
only in the choice of penetrating cryoprotectant, held at 12\% vol/vol)
we use the multi-solute osmotic virial equation to calculate the
chemical potential of the liquid, and its relation to calculate the
melting point (\emph{see} \textbf{Section 2.1}). Solution recipe
and all virial coefficients are shown in \textbf{Table 1} (note that we
approximate the 3-O-methyl-glucose employed as glucose monohydrate, as
thermophysical properties of the former are not available). Electrolytes
for which virial coefficients are not available are treated as ideal. We
calculated the unfrozen solution osmolalities and melting points to be
\(\pi_{0}\) = 3.16 Osmol/kg and --5.9 \textdegree C for propylene glycol, be
\(\pi_{0}\) = 3.07 Osmol/kg and --5.7 \textdegree C for glycerol, and \(\pi_{0}\) =
3.8 Osmol/kg and --7.1 \textdegree C for ethylene glycol.

\begin{table}[t]
\caption{Partial freezing solution and corresponding osmotic virial parameters. Solution composition from Tessier et al.~\cite{r1} and virial parameters from Zelinski et al.~\cite{r89} (PEG-35,000 virial coefficient from Wang et al.~\cite{r90}).}
\label{tab:virialparams}
\centering
\small
\setlength{\tabcolsep}{4pt}
\begin{tabularx}{\textwidth}{@{}>{\raggedright\arraybackslash}X c c c r r r@{}}
\toprule
\textbf{Component} & $\mathbf{k}_{\mathrm{diss}}$ & $\mathbf{B}_i$ & $\mathbf{C}_i$ &
\textbf{g/L} & \textbf{mM} & \textbf{mol/kg} \\
\midrule
HES      & 1 & 35.29  & $-1.107 \times 10^{7}$ & 50.00 & 0.25  & $3.38 \times 10^{-4}$ \\
Lactobionic Acid    & 1 & 0      & 0                     & 35.83 & 100   & 0.135 \\
Potassium Phosphate monobasic& 2 & 0      & 0                     & 3.40  & 25    & 0.0338 \\
Magnesium Sulfate (heptahydrate) & 2 & 0  & 0                     & 1.23  & 5     & 0.00674 \\
Raffinose (pentahydrate)     & 1 & 0.08   & 0                     & 17.83 & 30    & 0.0405 \\
Potassium Hydroxide          & 1 & 0      & 0                     & 5.61  & 100   & 0.135 \\
Sodium Hydroxide, 5N         & 1 & 0      & 0                     & 1.00  & 25    & 0.0338 \\
PEG-35,000   & 1 & 5200   & 0                     & 50.00 & 1.4   & 0.00193 \\
Trehalose (dihydrate)  & 1 & 0.12   & 0                     & 18.92 & 50    & 0.0676 \\
Glucose (monohydrate)  & 1 & 0.044  & 0                     & 19.42 & 98    & 0.132 \\
Propylene glycol   & 1 & 0.039  & 0                     & 124.8 & 1640  & 2.22 \\
\bottomrule
\end{tabularx}
\end{table}

In \textbf{Fig. 12}, for each choice of penetrating cryoprotectant, we
show the fraction of the system that freezes, the total solute
concentration in the remaining unfrozen liquid, and the individual
concentrations of each component as functions of temperature at
thermodynamic equilibrium. Several salient insights emerge, which we may
evaluate against the observations of Tessier et al. in their
histological and functional evaluations of partially frozen rodent
livers.

They first investigated differences between the three cryoprotectants at
--10 \textdegree C, observing glycerol to be broadly damaging across metrics, and
the two glycols to be less so (though in different respects). Regardless
of the known osmotic drawbacks of glycerol attributable to its slow
penetration, these findings appear generally predictable based on
\textbf{Fig. 12}, which shows that glycerol both allows the most ice
formation and drives the largest electrolyte imbalances in the remaining
liquid phase (e.g. \textasciitilde178 mM potassium) at --10 \textdegree C. Amongst
ethylene glycol and propylene glycol, the authors observed marginally
superior structural preservation (in histology) for ethylene glycol,
consistent with its relatively lower ice fraction, but marginally worse
functional outcomes in some cases nonetheless, which may be attributable
to mild differences in biochemical toxicity or osmotic stress profile of
the CPA at sub-0 \textdegree C temperatures, or an insufficient normothermic
perfusion timeline to observe complete recovery, which has been found in
select other models to take longer than one hour after prolonged sub-\textdegree C
storage \cite{r91}.

The authors proceeded to investigate propylene glycol at --10 \textdegree C and --15
\textdegree C, finding that storage at --15 \textdegree C, despite the putative benefit in
metabolism reduction (\emph{see} \textbf{Fig. 15d} in Case Study 2),
yielded more extensive hepatocellular injury, loss of liver sinusoidal
endothelial cells, and ultimately poorer functional outcomes. Per
\textbf{Fig. 12a}, this additional morphological and endothelial damage
is likely a consequence of the nearly 50\% increase in ice fraction
(from \textasciitilde0.3 to \textasciitilde0.44) upon shifting from --10
\textdegree C to --15 \textdegree C. Intriguingly, the total intercellular space in the liver
has been measured as \textasciitilde30-40\% of the total liver volume
\cite{r92}, which may further explain the relative palatability
of partial freezing at --10 \textdegree C, though it is unclear whether functional
degradation is driven more by mechanical aspects of ice formation, or by
the rate and degree of electrolyte ripening in the remaining liquid
phase (which is of course coupled to said ice). We note that in later
work \cite{r84}, the same authors saw somewhat improved outcomes
at --15 \textdegree C upon incorporating more polyethylene glycol, which may serve
to retard both ice growth and the diffusion of rejected electrolytes,
thereby reducing the sharpness of mechanical or osmotic shock
experienced.

These simple analyses both help to interpret past experimental outcomes
and suggest design approaches to for future partial freezing protocols.
For example, by altering the electrolyte concentrations in the starting
media, a solution could easily be designed which achieves the desired
sodium/potassium balance \emph{only at sub-0 \textdegree C} equilibrium, rather
than allowing the significant excursions seen here. Specific ice growth
limits may also be easily targeted; different candidate cryoprotectants
may easily be screened; and etc.

\subsection{Kinetic techniques}

The techniques described in the preceding section are unified by their
concern with achieving or maintaining stable thermodynamic equilibrium,
wherefrom additional spontaneous evolution of the system (e.g. phase
change, compositional change) becomes impossible. As such, from a
thermophysical perspective, these techniques are driven by phenomena
with long-to-infinite associated time scales, and depend only minimally
on acutely transient processes. However, many cryopreservation protocols
may deviate significantly from this equilibrium, passing through (or
deliberately occupying) a metastable supercooled state wherefrom ice
nucleation, whether desirable or to be avoided, is strongly \emph{time}
dependent, and for which temporal aspects of the protocol are key. In
this sub-section, we will review several emerging techniques dominated
by kinetic aspects of ice nucleation and growth, seeking to interpret
each within the kinetic framework described in \textbf{Section
2.2}.

\subsubsection{Ice seeding}

Kinetic techniques in general seek either to \emph{accelerate} or
\emph{delay} the nucleation of ice upon cooling beneath the melting
point. The process of \emph{accelerating} nucleation is often referred
to as ``ice seeding'', and is typically desired for two reasons: 1) to
avoid the excessive mechanical, thermal, and osmotic damage that can
occur if extracellular nucleation occurs at deeply supercooled
temperatures, and 2) to avoid intracellular ice nucleation entirely, by
enabling extracellular ice growth and solute rejection to gradually draw
water out of (and force cryoprotectants into) the cell, thereby
stabilizing it against nucleation as it reaches ever-colder
temperatures.

As such, within the formalism of nucleation theory, ice seeding
protocols must be designed such that the induction time of ice
nucleation in the extracellular space is much shorter than that in the
intracellular space. While the relative volume of a given cell compared
to the extracellular space already favorably biases these induction
times, this bias is typically insufficient to avoid the mechano-osmotic
damage that can be associated with rapid extracellular ice growth. As
such, emergent approaches to ice seeding focus on means to further shift
the induction time of extracellular nucleation forward, using both
passive approaches (e.g. ice nucleating agents (INAs)) or active
approaches (e.g. ultrasonication). Many excellent works over the past
decade have reviewed both the fundamentals of and progress in ice
seeding \cite{r93,r94,r95}; as such we will here review only the most
recent advances in the field.

Passive methods of ice seeding focus on increasingly effective ice
nucleating agents, which leverage heterogeneous nucleation modes to
induct ice formation at temperatures ever-closer to the equilibrium
melting point of the solution. Historically, the \emph{SnoMax} ice
nucleating bacteria, silver iodide, and a small clutch of other
nucleators were used to this end, but the past half decade has seen a
rapid expansion in INA offerings. Key materials of recent interest
include: 1) several varieties of naturally-occurring mineral, including
various feldspars, quartz, and the ``hyperactive'' K-feldspar LDH1,
which has been found to enable nucleation at \textless1 \textdegree C supercooling
\cite{r96}; 2) emergent classes of nature-derived water soluble
organic materials, including polysaccharides from the pollen
\emph{Carpinus betulus} (or European hornbeam) {[}97{]}, hydrated
cholesterol \cite{r98}, etc.; and 3) ice nucleating bacteria,
including previously unrecognized members of the Pseudomonadaceae,
Enterobacteriaceae, Xanthomonadaceae and Bacillaceae families
\cite{r99}. These molecules nucleate ice through a wide host of
divergent hypothesized mechanisms, from crystalline templating /
interfacial free energy reduction to specific bacterial protein
interactions, significant headroom remains in our understanding of
passive ice nucleation processes.

Active methods meanwhile do not involve additives to solution. One
popular such method in recent years is nucleation by acoustic or
ultrasonic stimulation, which is generally understood to drive
cavitation effects that yield massive hyper-local fluctuations in
pressure that produce ephemeral nuclei of high-pressure ice phases,
which then seed hexagonal ice \cite{r100}. Ultrasonic ice seeding
has shown beneficial results in the preservation of samples ranging from
hepatocytes to fresh food products \cite{r101,r102,r103}, though
significant additional work is needed to develop a physics-first
understanding of optimal sonication parameters for nucleation. As
evidence, it should be noted that ultrasound is also drawing
considerable recent attention as a means of supplying rapid warming for
large biological samples, wherein the goal is to \emph{avoid} ice
formation \emph{en route} from the vitrified state \cite{r103}.

\subsubsection{Supercooling}

As we have established throughout this chapter, the induction of ice
nucleation upon cooling beneath a solution's melting point occurs over a
finite and probabilistically quantifiable time scale. If this time scale
is greater than the desired preservation time for a given experiment,
the solution may remain at metastable equilibrium in a supercooled
liquid state, thereby enabling ice-free preservation at sub-0 \textdegree C,
sub-melting point temperatures.

Supercooling, though recently reaching new heights of popularity as
means of practical stabilization have expanded, is one of the
foundational concepts of cryopreservation for complex biological
systems, seen both in the literature for \textgreater60 years and in
nature for millennia. Supercooling as a fundamental mode of
\emph{self}-preservation in nature has been corroborated time and again
across organismal complexities, from single-celled fungi (e.g. yeast,
demonstrated by Franks and colleagues to stably supercool at --20 \textdegree C for
\textgreater16 weeks \cite{r104}) to whole mammals (e.g. the
arctic ground squirrel, demonstrated to supercool at \textasciitilde{}
--3 \textdegree C for months of winter on end \cite{r105}), and has inspired
the recent advent of supercooled cell \cite{r106}, tissue
\cite{r107}, and organ
preservation \cite{r3,r75,r76,r77,r108,r109}.

\begin{figure}[h]
\centering
\includegraphics[width=0.8\textwidth]{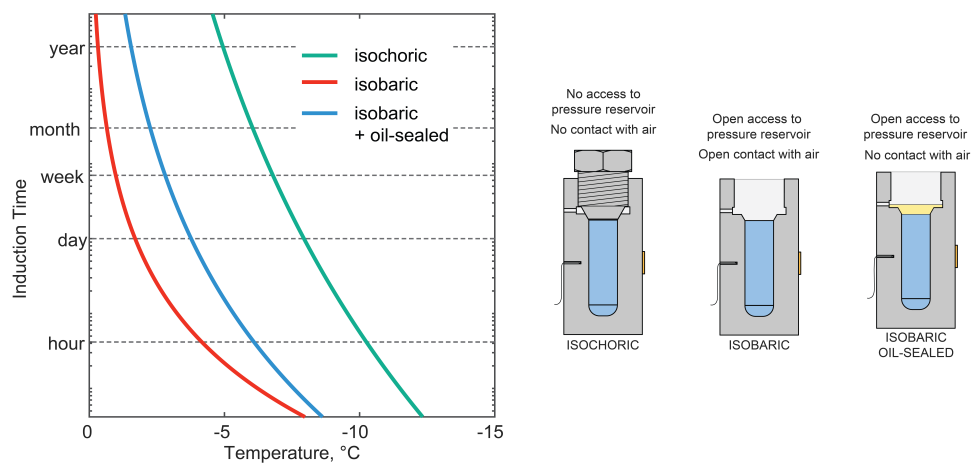}
\caption{Nucleation induction times as a function of temperature
for isochoric (green), isobaric (red), and oil surface sealed isobaric (blue) systems (left) and corresponding schematics of systems (right). Adapted with permission from Consiglio et al. \cite{r40}.}
\label{fig:fig13}
\end{figure}

\begin{figure}[h]
\centering
\includegraphics[width=1.0\textwidth]{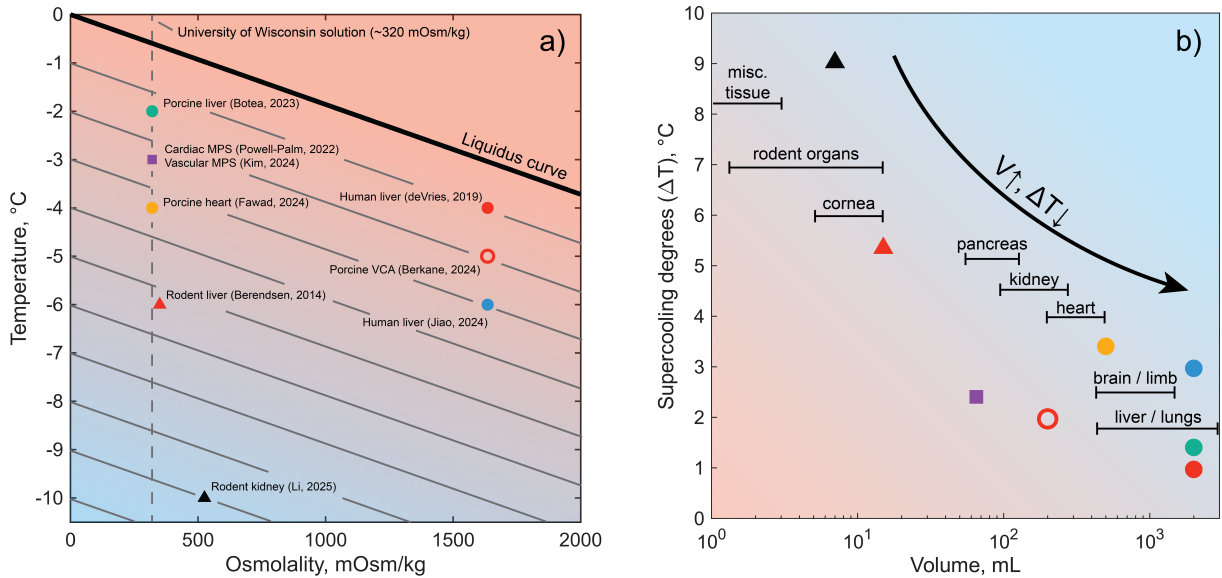}
\caption{Recent progress in supercooling for organ preservation.
a) Supercooling temperatures reported for organ and organoid models as a function of the osmolality of the solution used \cite{r3,r75,r76,r77,r91,r107,r108,r109,r115}. Thin gray lines represent supercooling isotherms, along which the distance in temperature from the melting point is constant. b) Degrees of supercooling past the melting point (i.e., \(\mathrm{\Delta}T\)) for the same solutions as a function of sample volume. The volume ranges corresponding to various human organs and tissues of interest in transplantation are indicated by horizontal black bars. The exponentially inverse correlation between sample volume and
supercoolability is consistent with classical nucleation theory and the extreme value interpretation of nucleation statistics \cite{r32,r122,r123}.}
\label{fig:fig14}
\end{figure}

From a physical perspective, the principal challenge of supercooling is
ensuring the \emph{stability} of the supercooled state, or the
probability that nucleation will occur (i.e. the induction time will be
reached) within the desired preservation period. In the modern era,
select intersecting approaches to supercooling stabilization have
emerged in biopreservation contexts, though many more may be imagined
based on the principles described in this chapter.

Usta and colleagues discovered that the liquid-air interface contributes
in a uniquely potent fashion to heterogeneous ice nucleation in
supercooled solutions, and that this contribution may be reduced in
large part by sealing the exposed surface of the solution with an
immiscible layer of liquid (e.g. mineral oil). This principle has since
been deployed extensively in the supercooling of various sensitive cell
models \cite{r106,r110,r111,r112,r113,r114}, and to a lesser degree in
preservation of mammalian livers \cite{r3,r75} and kidneys
\cite{r115} (wherein chemical stabilization is the dominant
driver; see below).

Powell-Palm and colleagues then identified that isochoric or confined
conditions (embodied by sealing the system of interest in a rigid,
air-free container) further stabilized the supercooled state
\cite{r40}, driven by a suite of non-chemical mechanisms
including the air-avoidance identified by Usta \cite{r116},
isolation against vibration \cite{r117,r118} and density
fluctuations (which are necessarily unconstrained in open systems with
access to atmospheric pressure), and putative atomistic effects on the
nucleation barrier \cite{r119}. A comparison of nucleation
induction times in pure water for both these techniques (and an open,
unaffected sample) are shown in \textbf{Fig. 13} (adapted from Consiglio
et al. \cite{r40}). Isochoric supercooling has since been
deployed in the preservation of cells, engineered tissue constructs,
mammalian livers, and mammalian hearts, in addition to a variety of
fresh food products \cite{r19}.

It is critical to note that the two mechanisms above, oil sealing and
isochoric supercooling, present \emph{purely physical, non-chemical}
means of stabilizing the supercooled state---i.e. they do not require
that any particular chemical composition be introduced into the
supercooled biologic. However, while this non-chemical nature is no
doubt favorable from a regulatory perspective (and enables interrogation
of purely temperature-dependent biopreservation phenomena, absent ice
and absent chemical confounders), nothing \emph{precludes} these
techniques from being \emph{coupled} with chemical interventions to
further stabilize supercooling, as noted previously by Taylor et al.
\cite{r74}.

An initial demonstration of this combination was provided by the group
of Toner and Uygun, who coupled oil-sealing with penetrating
cryoprotectants and high-molecular weight polymers to achieve multiday
supercooled preservation of the human liver, providing a watershed
moment in the demonstration of the applicability of the technique to
transplant medicine. However, despite this blend of stabilizing effects,
supercooling was only achieved to \textasciitilde1 \textdegree C beneath the
melting point (--4 \textdegree C storage temperature) \cite{r3}. More
recently, Huang and colleagues have realized this premise to a much
greater degree, obtaining supercooling to \textasciitilde9 \textdegree C beneath
the melting point (--10 \textdegree C storage) in rodent kidneys \cite{r115}.
To achieve this, they combined oil-sealing with a preservation solution
that includes the synthetic ice blocker 1,4-cyclohexanediol alongside
multiple polymers, each at molecular weights at which they will both
leave the vascular space (i.e. enter the interstitium) and plausibly
provide molarity-disproportionate depression of the nucleation
temperature \cite{r120,r121}. Huang and colleagues note, as we
have also previously \cite{r19}, that even deeper supercooling
may likely be achievable by combining this chemical approach with
isochoric supercooling. A survey of recent organ supercooling studies is
shown in \textbf{Fig. 14}.

The practical utility of supercooling, especially for large,
ultra-high-value biologics like transplantable human organs, is a strict
function of the likelihood of ice formation at the desired storage
temperature, in the precise configuration to be used. While for many
years the probabilistic nature of ice nucleation had been considered an
insurmountable difficulty in achievement of a reliable technique, recent
work (borrowing from long-standing insights in metallurgy and
phase-change thermal energy storage) has leveraged the robust
mathematical describability of memoryless activated stochastic processes
to powerfully constrain the nucleation probability space, enabling
rational design of \emph{safe, stable} supercooling protocols by
leveraging the very stochasticity that previously appeared a liability.
Using constant-cooling-rate or isothermal nucleation experiments, the
theory described in \textbf{Section 2.2.5} of this chapter may
be used to calculate an induction time isopleth \emph{for a given
probability of ice} nucleation (say, 1-in-10,000), thereby enabling
quantification, from an empirical, protocol- and system-specific basis,
of a safe supercooling duration. In Case Study 2, we detail this
thermodynamic-kinetic approach to ``supercooling engineering''.

\subsubsection{Case study 2: Rational engineering design of supercooling
protocols}

Using the statistical nucleation kinetics framework described in
\textbf{Section 2.2.5}, benchtop nucleation experiments may be
used to rigorously design supercooling protocols with arbitrarily high
margins of safety against nucleation, or arbitrarily low probabilities
of ice nucleation at a desired temperature or a desired storage period.
The general process is as follows:

1. For the specific container type, sample volume, and sample
composition of interest, perform multiple constant cooling rate
nucleation temperature tests in order to assemble a distribution of
nucleation temperatures (a ``survivor curve'' or cumulative
distribution). A minimum of \textasciitilde30 replicates is recommended
to robustly capture intra-sample variance (i.e., using \emph{precisely}
the same sample repeatedly), a process that can be informed by computing
confidence intervals (e.g., through bootstrapping).

a) \emph{Note: If these replication numbers prove untenable at full
sample volume due to cooling rate or other limitations, scaled-down
versions of the desired system may be used, and surface area or
volume-based scaling rules may be applied to adjust the results as
needed} \cite{r32}\emph{.}

\emph{2.} Fit a non-homogeneous Poisson distribution (Equation 47) to
this data in order to extract the temperature-dependent system
nucleation rate \cite{r40}\emph{.} Use this rate to calculate the
probability of ice nucleation at desired storage time or/and the
induction time for a desired probability of nucleation as functions of
temperature.

3. Repeat steps 1 and 2 with different samples in order to capture
inter-sample variance. Take worst case trial (most nucleation prone),
average trial, or apply extreme value statistical methods in order to
compute the overall freezing probability \cite{r32}.

4. Identify a combination of storage temperature and storage time that
provide the desired margin of safety on supercooling stability (e.g., a
1-in-10,000 chance of ice nucleation). For any given physical
system/container with the same heterogeneous nucleation characteristics
(i.e. same surfaces in contact with the liquid, same closure conditions,
etc.) as captured in the underlying nucleation data from Step 1,
extrapolate these results in sample volume, surface area, or solution
composition using available physical and thermodynamic scaling rules to
further explore the opportunity space.

Using data from Consiglio et al., we show sample outputs of this process
in \textbf{Fig. 15} (for a single sample). We use the preferred form
\(J = A\exp\left( - \frac{B^{'}T_{\mathrm{m}}^{5}}{T^{3}\Delta T^{2}} \right)\)
to capture the temperature dependence of the nucleation rate, with
parameters \(A = 2.01\), \(B' = 1.07\), representing data for 1 L pure
water contained within an isochoric container, and in direct contact
only with a uniform petrolatum coating on the chamber walls (\emph{see}
Fig. 5 of Consiglio et al. \cite{r32}). The freezing probability
is calculated as \(p = 1 - e^{- Jt}\).

In \textbf{Fig. 15a}, we show the probability of freezing in a 7-day
period as a function of solution melting point. Here, the
\(T_{\mathrm{m}} = 0\ \)\textdegree C curve is calculated according to the steps above from
the benchtop data described, and the \(T_{\mathrm{m}} < 0\ \)\textdegree C data utilize the
assumption that, in the dilute limit, the nucleation parameters \(A\)
and \(B\) are insensitive to the concentration of conventional
permeating CPAs. This assumption reproduces the common observation that
the ratio \(\lambda\) of nucleation temperature depression to melting
temperature depression is equal to 1 for solutions at bulk volumes
dominated by heterogeneous nucleation \cite{r40,r124}. We note
however that \(\lambda\ \sim\ 2\) in the same solutions when in
emulsified or atmospheric droplet forms susceptible to homogeneous
nucleation. Furthermore, solutions with polymeric additives may reach
\(\lambda \gg \ 2\) (whether heterogeneously or homogeneously
nucleating), a phenomenon which Powell-Palm et al. have attributed to
the dominant role of the configurational entropy of the liquid in
prescribing the liquid-ice interfacial free energy \cite{r9}.

Using the same data, in \textbf{Fig. 15b}, we show the probability of
freezing as a function of degree of supercooling
\(\mathrm{\Delta}T = T_{\mathrm{melting}} - T_{\mathrm{supercooling}}\). In \textbf{Fig.
15c} we show the temperatures to which four common binary
water-cryoprotectant solutions are predicted to stably supercool for 1
week with a 1-in-1,000 chance of ice nucleation. Equilibrium melting
temperature curves are calculated via the osmotic virial equation
(\emph{see} \textbf{Section 2.1}).

\begin{figure}[t]
\centering
\includegraphics[width=0.8\textwidth]{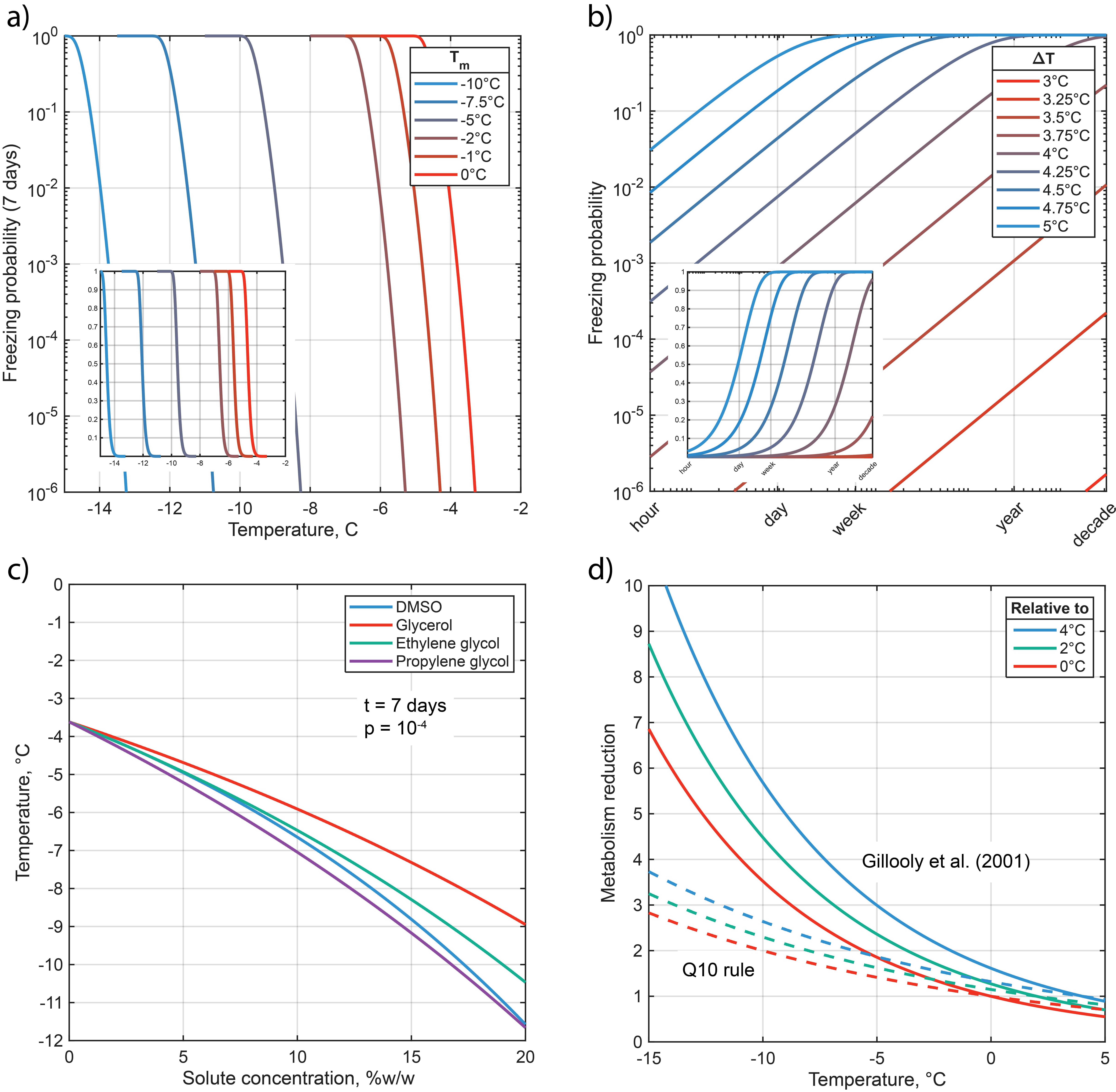}
\caption{Kinetics insights for the rational design of safe
supercooling protocols. Panels and methodology described in detail in
the text above. a) Freezing probability for 7 days of supercooling at a given temperature, for solutions of varying melting temperatures. b) Temperature to which a solution may be supercooled for 7 days with a 1-in-1000 chance of freezing, for aqueous binary solutions of four common cryoprotectants. c) Freezing probability versus time for varying degrees of supercooling past the melting point. d) Factor of metabolism reduction relative to various conventional cold storage temperatures as a function of supercooling temperature. Dashed lines represent the oft-cited (though woefully poorly supported) notion that metabolism is reduced by half for every 10 \textdegree C of cooling. Solid lines represent the
landmark empirical metabolism model of Gilooly et al. \cite{r41}.}
\label{fig:fig15}
\end{figure}

Finally, in \textbf{Fig. 15d}, we show an additional essential
consideration in choosing a desired supercooling temperature - the
degree of metabolic suppression putatively achieved at that temperature.
Here, dashed lines show the factor by which metabolism is suppressed
relative to \(\geq\)0 \textdegree C classical storage temperatures according to the
oft-cited ``Q10 rule'', which supposes that the rate of metabolism
halves for every 10 \textdegree C of cooling from normothermia. This ``rule'' is
based simply on the requirement of Arrhenius scaling for the biochemical
reactions of which metabolism is comprised, and not life-specific
theoretical or empirical data. As a data-driven alternative, we also
provide the reduction in metabolism predicted for mammals by the
landmark empirical model of Gilooly et al. \cite{r41}. While
their model is based on data between 0 \textdegree C and 37 \textdegree C, we argue that this
scaling provides much better agreement with the last decade of
preservation extension observed in studies on supercooling of complex
biological systems.

In total, \textbf{Fig. 15} conveys the rich variety of physicochemical
information that may be used to rationally design safe and reliable
supercooling protocols, requiring only straightforward benchtop
nucleation temperature experiments and elementary statistical kinetics calculations.

\subsubsection{High pressure freezing}

High-pressure freezing (HPF) exploits the strong pressure-dependence of
water's phase diagram to kinetically suppress ice formation during
cooling, enabling vitrification or near-vitrification with substantially
lower cryoprotectant concentrations than are required at atmospheric
pressure. Raising pressure shifts the liquid--ice coexistence line and
reduces the driving force for nucleation and growth at a given
temperature.

In conventional HPF devices adapted from electron microscopy, small
specimens (typically $\leq$200 µm thick) are rapidly pressurized to
\textasciitilde200 MPa and simultaneously cooled by a high-velocity
liquid nitrogen jet or similar. Recent work shows that such platforms
can be repurposed from cryo-fixation to cryopreservation
\cite{r125}, with 2D monolayers and 3D spheroids vitrified at
\textasciitilde210 MPa with only 20--30\% v/v permeating CPA plus
non-penetrating polymers displaying higher post-thaw viability,
metabolic activity, and retention of cell-cell junctions than comparable
normal-pressure plunge-freezing protocols.

Closely related ``self-pressurized rapid freezing'' (SPRF) approaches
achieve comparable pressures without an external pressurization system
by rapidly cooling sealed capillaries or microchannels \cite{r126,r127} completely filled with aqueous media, absent air. Initial ice
expansion in the confined volume raises internal pressure into the 10s
to 100s MPa range, and the system follows an effectively isochoric path
in which further ice growth is disfavored. HPF and SPRF thus sit at the
intersection of equilibrium and kinetic considerations, where pressure
first reshapes the liquidus surface, then consequently weakens the
thermodynamic driving force for ice formation.

\subsection{Transport-driven techniques}

As we have argued thus far, cryopreservation, as a process dependent on
the management of phase change, is driven principally by equilibrium
thermodynamics and nucleation and growth. While these phenomena
obviously intersect with and are dependent upon heat and mass transport
in many critical ways, techniques truly \emph{driven} by transport
phenomena (i.e., moreso dependent on transport than on the underlying
thermodynamics and kinetics at play) are comparatively rare. In this
sub-section, we will review two such techniques.

\subsubsection{Directional freezing}

Directional freezing consists of advancing a biologic across a known
(typically linear) temperature gradient in order to control the speed
and position of the freeze front and ensure uniformity of freezing
conditions. Introduced first by Rubinsky \& Ikeda \cite{r128} as
an analytical tool with which to investigate ice front geometry and
cell-ice interactions, the technique has since been applied as a means
of minimizing damage associated with ice growth during preservation of
biological samples across scales.

Directional freezing devices typically utilize two constant-temperature
heat sinks, one at a temperature above the melting point of the
biological sample of interest and the other below, separated by a gap of
known length. A conducting substrate of low thermal mass is then placed
across sinks, such that the portion spanning the gap assumes a constant,
approximately linear temperature gradient between the two. A biological
sample in solution is then placed in intimate thermal contact with this
plate, nominally fixing its temperature to its position in the direction
of the gradient, and producing a freeze front perpendicular to this
direction at the position corresponding to the melting point. The sample
is then slowly advanced from the warm side to the cold, such that its
cooling rate is a function only of its linear velocity (see schematic in
\textbf{Fig. 16}).

By then controlling this sample velocity, directional freezing fixes the
local cooling rate at the moving freeze front, converting what is
ordinarily a spatially and kinetic heterogeneous freezing process into a
nigh-deterministic moving-boundary problem, describable by the classical
Stefan formulation \cite{r129}. Furthermore, given that latent
heat and rejected solutes may both be evacuated in a prescribed
direction (into the unfrozen portion), multi-site nucleation and
re-melting related to recalescence events are suppressed, and solute
concentration gradients remain approximately constant at the freeze
front during freezing. In principle, directional freezing thus enables
freezing of an entire biological sample at a uniform cooling rate,
solute-rejection rate, etc.

In practice, this approach has been productively utilized in a variety
of samples for which the optimal aspect ratio for enforcement of the
linear temperature gradient by conduction (footprint dimensions >> thickness) can be obtained. For adherent
cellular monolayers, Braslavsky and colleagues \cite{r130} have
shown that a two-step implementation incorporating an initial
directional freeze to --20 \textdegree C followed by conventional slow cooling to
cryogenic temperatures has been shown to markedly improve post-thaw
attachment and viability relative to conventional slow cooling alone.
Others have shown similar benefit in the freezing of spermatozoa across
multiple species and multiple cell suspensions \cite{r131,r132,r133,r134},
where uniform front propagation mitigates localized crystallization and
solute pocketing that otherwise accompany uncontrolled freezing.

Directional freezing has also been investigated to a very limited degree
at tissue and organ length scales, for which increasingly sophisticated
stages involving multiple temperature gradients have been developed
\cite{r135}. Notably, Arav and co-workers have demonstrated
improvement in frozen preservation of both ovarian tissue slices
\cite{r136,r137,r138} and whole sheep ovaries, some of which proceeded
to produce hormonal activity upon thawing and transplantation. However,
significant additional research effort is needed to establish potential
utility of the technique for organ-scale biologics, and scaling analysis
by Ukpai \& Rubinsky suggests that samples above \textasciitilde1 cm
characteristic thickness cannot be directionally frozen in a uniform
fashion with the standard single-conductive-substrate configuration
\cite{r129}.

We also note that directional freezing presents a purely physical tool
for modulation of freezing behaviors, and thus may be coupled with
related chemical techniques, producing for example ``directional partial
freezing'' or ``directional isochoric freezing''.

\begin{figure}[h]
\centering
\includegraphics[width=1.0\textwidth]{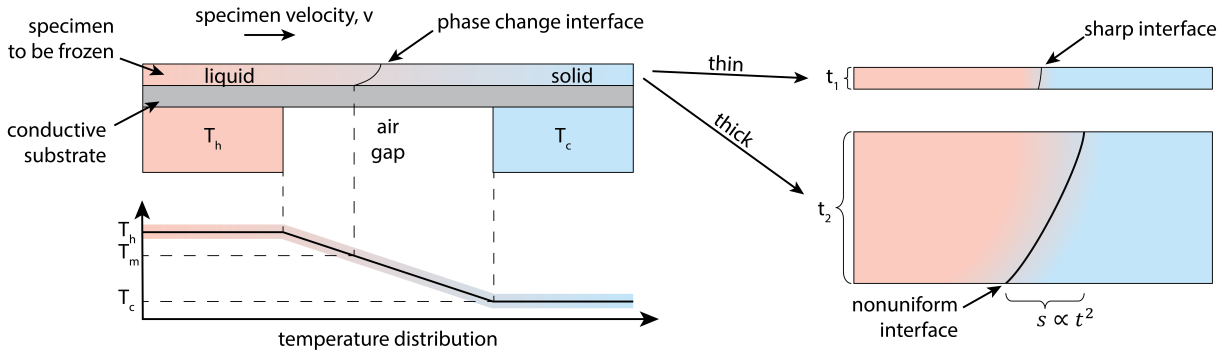}
\caption{Illustration of a typical directional freezing
platform. A biological sample is advanced along a conductive substrate with an imposed (typically linear) temperature gradient, such that its velocity controls the rate at which the liquid-ice phase change interface moves through the sample (i.e. the freezing rate). As demonstrated by Ukpai \& Rubinsky \cite{r129}, while this
interface is often assumed ``flat'', or perpendicular to the direction of the temperature gradient, it is only so for extremely thin samples, and scaling analysis suggests that uniform directional freezing of biologics with greater than \textasciitilde1 cm thickness cannot be achieved using conventional single-plate conduction in a room temperature atmosphere.}
\label{fig:fig16}
\end{figure}

\subsubsection{Spray, droplet, and flash freezing}

Since Birdseye's foundational discovery in the 1920s that faster cooling
produces smaller ice crystals, rapid-cooling approaches have become
integral to both microscale preservation of cellular products and
macroscale preservation of fresh food products \cite{r139,r140,r141,r142}.
In industrial contexts, a broad class of high--heat-flux freezing
technologies--- including cryogenic freezers, impingement freezers
\cite{r143}, fluidized bed freezers, and immersion freezers
\cite{r144}--- leverage large convective coefficients (up to
\textasciitilde2000 W/m\textsuperscript{2}), thin thermal boundary
layers, and varied sub-0 \textdegree C refrigerants (e.g., liquid or high-velocity
gaseous nitrogen, air at --40 \textdegree C, ethanol mixtures, etc.) to achieve high
cooling rates in mL- to L-scale food products, thereby maintaining
post-thaw quality by reducing characteristic crystal size, drip loss,
and textural degradation \cite{r140}. Intriguingly, while such
macroscale rapid cooling platforms are omnipresent in the food
preservation domain, they have not yet been widely applied to
rate-dependent whole organ or large organism cryopreservation efforts
(see vitrification section below), suggesting potentially low-hanging
opportunities for translational engineering efforts in the near future.
As noted in the excellent review by James et al. {[}144{]}, this may be
a consequence in part of knowledge in state-of-the-art food rapid
freezing technologies being disseminated through industrial pathways,
rather than through conventional scientific publishing.

At smaller scales, droplet- and spray-based rapid freezing methods
pursue the same transport objective by reducing the characteristic
thermal length scale. Cell-laden pL- to uL-scale droplets produced by
atomizing nozzles, microfluidic droplet generators, or inkjet heads,
when deposited on cryogenic substrates or into cryogenic liquids, can
achieve cooling rates in excess of $10^{4}$--$10^{6}$ \textdegree C/min,
sufficient to restrict crystalline ice growth to within tolerable limits
even with little-to-no penetrating cryoprotectants
\cite{r145,r146,r147,r148}. For a more in-depth discussion of droplet-scale cryopreservation approaches, see the Chapter of this book authored by Zhao and colleagues.

\subsection{Vitrification}

At long last, we reach that technique which above all others defies
clean placement in one or the other of the preceding categories:
vitrification. First explored for cryopreservation purposes by Luyet \&
Hodapp in 1938 \cite{r149}, vitrification describes the process
of shuttling an ice-free biological sample into and out of the glassy
state, and represents the most thermophysically complex cryopreservation
process yet devised. Successful practice of this technique requires an
extremely delicate balance between thermodynamic, kinetic, and transport
considerations, all of which grow exponentially more challenging as the
biological sample grows in size, and further requires reckoning with the
glass transition itself, a unique multi-physics phenomenon which, at the
most fundamental levels, remains mysterious \cite{r150,r151,r152}.
While a complete treatment of the theory and practice of vitrification
is outside the scope of this Chapter, we will here provide some salient
commentary on the current state and future prospects of vitrification. We also refer the reader to several recent reviews on vitrification \cite{r73,r153,r154}, and to the various other Chapters of this book that address it in varying capacities (e.g., the Chapter by Shu and colleagues, discussing the role of rewarming in vitrification, and the Chapter by Elliott and colleagues, relaying a technique for vitrification of articular cartilage).

\subsubsection*{Interplay between thermodynamics, kinetics, and transport}

To achieve truly ice-free vitrification of a biological sample, the
sample must travel through the metastable supercooled regime beneath its
melting point and reach the safety of the glass transition temperature
(at which dynamical arrest eliminates the possibility of phase change on
human time scales) in a period less than the induction time of ice
nucleation in the sample. As such, the success of vitrification in a
given sample is a complex function of solution thermodynamics, which
prescribe the distance in temperature that the sample must traverse
under threat of ice formation (between the melting point and glass
transition temperature); ice nucleation kinetics, which prescribe the
induction time of nucleation within this distance, as a function of
cooling rate; and transport, which prescribes the practitioner's ability
to achieve a cooling rate that enables the sample to reach the glass
transition before the according induction time. If the requirement of
\emph{total} ice avoidance is relaxed, ice \emph{growth} kinetics also
play a dominant role in this interplay, prescribing the degree to which
any allowed nanoscale nucleation will develop into deleterious micro- or
macro-scale ice formation. The basic logic of this interplay is laid out
in \textbf{Fig. 17}, and a review of thermodynamic aspects thereof is
also provided by Wowk \cite{r73}.

\begin{figure}[h]
\centering
\includegraphics[width=1.0\textwidth]{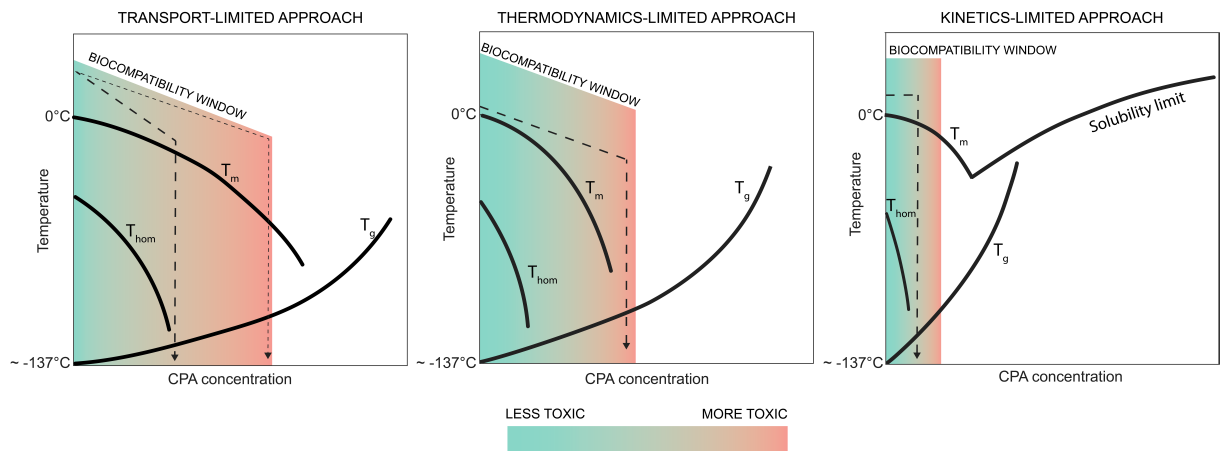}
\caption{In all panels, the biocompatibility window represents
the range of cryoprotectant concentrations over which vitrification
could be achieved with compromising the biologic through toxicity alone. The large-dashed lines represent a hypothetical
temperature-concentration trajectory for a sample. Tm represents the
ice-liquid melting curve; Thom represents the homogeneous ice nucleation temperature, under which rate-dependent avoidance of ice is assumed practically impossible; and \(T_{\mathrm{g}}\) represents the glass transition curve. Panel (a) shows the conventional transport-limited approach pioneered by Fahy and colleagues, wherein the sample is shuttled through the metastable region where ice formation is possible, but where it can be avoided by cooling and warming at sufficient speeds. Panel (b) shows a thermodynamics-limited approach, wherein novel cryoprotectants are utilized to bring the putative intersection of the melting and glass transition curves within the biocompatibility window, thereby enabling vitrification without entering a regime where ice nucleation is possible. Panel (c) shows a kinetics-limited approach, where high-molecular weight polymers or other molecules specifically suppressing the nucleation of ice (as opposed to its thermodynamic stability) are used to stabilize rate-independent passage through the metastable state. We note the role of solute solubility limits only in Panel (c), as they often prove trivially avoidable in multi-component molecular aqueous solutions, but much more challenging when polymers or large complex molecules are incorporated.}
\label{fig:fig17}
\end{figure}

\subsubsection*{Approaches to vitrification}

This multiphysics interplay obviates many levers by which to design a
successful vitrification process, and it is instructive, in
contemplating directions for the development of the field, to consider
each in the limit of ideal behavior.

The transport-limited ideal is by far the best explored in recent
literature, where it is well understood that in the limit of extremely
fast cooling and warming, sensitivities of the vitrification process to
the thermodynamic and kinetic parameters of the sample relax
dramatically (\textbf{Fig. 17}, left). This concept has been richly
demonstrated by Thorne and colleagues, who optimized a metallic sample
holder and cryogen exposure protocol for bovine oocytes in order to
achieve cooling rates up to 600,000 \textdegree C/min and warming rates up to
200,000 \textdegree C/min \cite{r155}; by Akiyama and colleagues, who
developed an inkjet-based droplet extrusion platform to achieve
intracellular vitrification in cells at cooling rates up to order
\textasciitilde1,000,000 \textdegree C/min and warming rates up to order
\textasciitilde10,000,000 \textdegree C/min \cite{r156}; by Bischof and
colleagues, who utilized low-thermal mass, high-thermal conductivity
meshes to achieve cooling and warming rates up to order
\textasciitilde100,000 \textdegree C/min in coral larvae, zebrafish embryos, and
other small organismal samples; and in many other works reviewed in
various recent papers \cite{r157}. Transport-driven vitrification
techniques have also gained recent popularity in the preservation of
large-scale biologics, with particular and extensive focus on new
electromagnetic and photothermal interventions during the rewarming
process, as surveyed in the Chapter of this book authored by Shu and colleagues.

As a brief aside, we note that curiously little effort has been applied
to maximizing \emph{cooling rates} of the same large-scale biologics, on
which, as recently pointed out by Thorne and colleagues
\cite{r158}, measured ``critical'' warming rates depend. This
oversight is likely a symptom of historical definitions of successful
vitrification, wherein some degree of ice formation (typically
\textasciitilde0.2\% by volume) has been accepted. In this case, wherein
potentially numerous ice nuclei are allowed to form on cooling, the
\emph{growth} of those nuclei becomes asymmetrical between cooling and
warming, with the vast majority occurring during warming. This asymmetry
in growth in turn gives rise to the perception that the ``critical''
thermal rates required for successful vitrification are much larger for
warming than for cooling. However, if successful vitrification is
instead defined as \emph{complete} avoidance of crystallization,
critical cooling rates and warming rates become \emph{precisely equal},
as ice nucleation is a memoryless stochastic process (i.e. agnostic to
the thermal direction in which the system is moving).

The thermodynamics-limited and kinetics-limited ideals for vitrification
have been investigated much more minimally. In the thermodynamic case,
we may envision a biocompatible cryoprotectant solution for which the
combination of melting point depression and glass transition temperature
elevation together shrink the temperature-width of the metastable zone
where ice nucleation is plausible, until safe passage through this zone
at achievable cooling/warming time scales becomes trivial (\textbf{Fig.
17}, middle). In the 1970s, this concept was powerfully demonstrated by
Elford \& Walter \cite{r69} in the cryopreservation and
functional revival of cm-scale smooth muscle tissues using DMSO as the
sole cryoprotectant and managing toxicity by liquidus tracking (we note
that while Elford \& Walter only stored the tissues at --80 \textdegree C, the
avoidance of ice for 24 hours at this temperature, in addition to the
concentration of DMSO used, suggest that equivalent storage beneath the
glass transition would also not result in freezing). However, in recent
literature, and despite the exploding recent interest in vitrification
of complex, multi-scale biologics, development towards (and even serious
discussion of) this thermodynamic ideal has largely vanished, in favor
of transport-limited strategies, pioneered by Fahy and colleagues, built
around achieving ``critical'' cooling and warming rates needed to avoid
substantial ice growth in marginally stable systems (\textbf{Fig. 17},
left). Notable exceptions to this absence are the review works of Taylor
and colleagues \cite{r74,r159} and the mid-2000s research by
Brockbank and colleagues on an ultra-concentrated (83\% cryoprotectant
by mass) derivative of the vitrification solutions developed by Fahy
\cite{r160,r161}. We suggest, especially in light of recent
progress in the discovery of new cryoprotectants that remain
biocompatible at thermodynamically-relevant concentrations
\cite{r162,r163}, that this thermodynamics-limited ideal approach
merits significant additional consideration.

Considering the past \textasciitilde15 years of advances in our
understanding of ice nucleation and growth processes in aqueous organic
solutions, a compelling kinetics-limited ideal may also be articulated,
wherein molecular or polymeric additives that interact with water and
ice in increasingly \emph{specific} ways may drive up the induction time
of nucleation or inhibit post-nucleation growth, enabling ice-free
cooling and warming at achievable rates in systems that retain high
equilibrium melting points (\textbf{Fig. 17}, right). Ice
recrystallization inhibitors, anti-freeze proteins, ice-modulating
polymers \cite{r164,r165}, and other agents affecting ice at the
critical or sub-critical cluster scale provide promising routes to this
end, especially in smaller-scale systems where the viscosities
associated with high-molecular weight polymers need not preclude their
use at high concentrations (as it may in vascular systems loaded by
perfusion). We note further that, given the dominance of heterogeneous
nucleation in vitrification of bulk (\(\geq\)mL-scale) systems, methods
developed to suppress heterogeneous ice nucleation during
\emph{supercooled preservation} may be applied directly to
vitrification, including elimination of air-liquid interfaces
\cite{r40,r117}, isolation from the atmospheric pressure
reservoir via isochoric conditions \cite{r40}, incorporation of ice-phobic surfaces, etc.

\subsubsection*{Toxicity, cracking, and the challenges in the vitrification of
increasingly large systems}

Each of the approaches described above has been demonstrated (to at
least a nominal degree) in vitrification of \emph{small} biological
samples (generally pL- to uL-scale, and up to \textasciitilde1 mL),
where the intrinsically low thermal diffusivity of aqueous organic
systems does not broadly prevent rapid cooling and warming;
less-than-millimeter mass diffusion length scales allow effective
delivery of thermodynamically-favorable cryoprotectants; and small
volumes and surface areas of water within the sample minimize nucleation
rates. However, in scaling to the many-mL to many-L systems relevant to
organ \cite{r49,r166}, whole-organism cryoconservation
\cite{r167,r168}, or biostasis for space travel
\cite{r169}, universal additional challenges emerge, including
solution biocompatibility at prolonged exposure times
\cite{r170}, resilience against thermomechanical stress and
fracture \cite{r51}, and intersecting general difficulties in
solution stability and deliverability.

While these problems remain far from solved, and successful vitrified
cryopreservation of human-sized organs, mature organisms, etc. remains
elusive, recent advances in \emph{materials discovery} (for
biocompatibility \cite{r78,r79,r163,r171}; resistance to cracking
\cite{r50}; profound interaction with water \cite{r16,r172,r173}; thermal rate optimization \cite{r174,r175}; viscosity
optimization \cite{r55}; etc.) offer a rapidly expanding
parameter space from which to formulate optimal new vitrification
protocols, according to \emph{any} of the design principles articulated
above. As such, we encourage the reader to approach their own work in
vitrification with an open mind, and with consideration of the rapidly
evolving physical and material bases of the field.

\section{Conclusion}
In this Chapter, we have sought to establish a self-consistent
understanding of the physical principles driving all cryopreservation
processes, in terms of a discrete set of accessible theories rooted
ultimately in Gibbsian thermodynamics. We hope that the reader will
carry from this work an increased appreciation of the fundamental
similarities and differences between emerging cryopreservation
technologies, and of the myriad opportunities to leverage these
similarities and differences to hybridize, optimize, and otherwise
improve their application of these technologies. We also hope that the
extreme mismatch between the potential for thermodynamic analysis in
cryopreservation and the degree of analysis that has been performed will
be evident, and that this mismatch will inspire ever-increasing effort
in the application of thermodynamics to the pressing, impactful problems
of cryobiology.

\section*{Acknowledgments}
The authors acknowledge the support of the National Science Foundation (NSF) Engineering Research Center for Advanced Technologies for Preservation of Biological Systems (ATP-Bio), NSF EEC \#1941543.

\end{document}